\documentclass[useAMS,usenatbib]{mn2e}
\usepackage{graphicx,fleqn,natbib}
\DeclareGraphicsExtensions{.eps,.cps,.ps,.eps.gz,.ps.gz,.eps.Z}
\usepackage{subfigure}

\title[A statistical study of UVOT GRB afterglows.]{A statistical study of gamma-ray burst afterglows measured by the {\it Swift} Ultra-violet Optical Telescope}

\author[Oates et al.]{S. R. Oates$^{1}$, M. J. Page$^{1}$, P. Schady$^{1}$, M. de Pasquale$^{1}$, T. S. Koch$^{2}$, A. A. Breeveld$^{1}$,
\newauthor P. J. Brown$^{2}$, M. M. Chester$^{2}$, S. T. Holland$^{3,4,5}$, E. A. Hoversten$^{2}$, N. P. M. Kuin$^{1}$,
\newauthor F. E. Marshall$^{3}$, P. W. A. Roming$^{2}$,  M. Still$^{1}$, D. E. Vanden Berk$^{2}$, S. Zane$^{1}$
\newauthor and J. A. Nousek$^{2}$\\
$^{1}$ Mullard Space Science Laboratory, University College London, Holmbury St. Mary, Dorking Surrey, RH5 6NT, UK; sro@mssl.ucl.ac.uk \\
$^{2}$ Department of Astronomy and Astrophysics, Pennsylvania State University, 104 Davey Laboratory, University Park, PA 16802 \\
$^{3}$ Astrophysics Science Division, Code 660.1, NASA Goddard Space Flight Centre, 8800 Greenbelt Road, Greenbelt, Maryland 20771, USA \\
$^{4}$ Universities Space Research Association, 10211 Wincopin Circle, Suite 500, Columbia, Maryland 21044, USA \\
$^{5}$ Centre for Research and Exploration in Space Science and Technology, Code 668.8, NASA Goddard Space Flight Centre,\\
8800 Greenbelt Road, Greenbelt, Maryland 20771, USA }

\begin{document}

\maketitle

\label{firstpage}

\begin{abstract} 
We present the first statistical analysis of 27 Ultra-violet Optical Telescope (UVOT) optical/ultra-violet lightcurves of 
Gamma-Ray Burst (GRB) afterglows. We have found, through analysis of the lightcurves in the observer's frame, that a 
significant fraction rise in the first 500s after the GRB trigger, that all lightcurves decay after 500s, typically 
as a power-law with a relatively narrow distribution of decay indices, and that the brightest optical afterglows 
tend to decay the quickest. We find that the rise could either be produced physically by the start of the 
forward shock, when the jet begins to plough into the external medium, or geometrically where an off-axis 
observer sees a rising lightcurve as an increasing amount of emission enters the observers line of sight, 
which occurs as the jet slows. We find that at 99.8\% confidence, there is a correlation, in 
the observed frame, between the apparent magnitude of the lightcurves at 400s and the rate of decay after 
500s. However, in the rest frame a Spearman Rank test shows only a weak correlation of low statistical significance between
luminosity and decay rate. A correlation should be expected if the afterglows were produced by off-axis jets, suggesting 
that the jet is viewed from within the half-opening angle $\theta$ or within a core of uniform energy density 
$\theta_c$. We also produced logarithmic luminosity distributions for three rest frame epochs. We find no evidence 
for bimodality in any of the distributions. Finally, we compare our sample of UVOT lightcurves with the X-ray 
Telescope (XRT) lightcurve canonical model. The range in decay indices seen in UVOT lightcurves at any epoch is 
most similar to the range in decay of the shallow decay segment of the XRT canonical model. However, in the 
XRT canonical model there is no indication of the rising behaviour observed in the UVOT lightcurves.

\end{abstract}


\section{Introduction}
\label{intro}

Gamma-ray bursts (GRBs) release between $10^{52}$ and $10^{54}$ ergs during the prompt emission, which 
lasts from a few milliseconds to a few thousand seconds, and is followed by an afterglow, which is observed 
in the X-ray to radio range from as little as a few tens of seconds up to several months after the GRB trigger. 

The energy is transported in a relativistic outflow \citep{mes97} that is likely anisotropic 
\citep{sar99} and the energy is expected to be released by internal and external shocks. Internal shocks 
\citep{rees94} are thought to produce the prompt gamma-ray emission, while external shocks 
\citep{rees92} are thought to produce the afterglow. Internal shocks occur when shells of material, 
which are thrown violently from the progenitor at different Lorentz factors, overtake each other. The 
external shocks are produced when the shells of material are decelerated by the external medium. 

The short duration of the gamma-ray emission and the rapid decay of the afterglow motivated the construction 
and launch of {\it Swift}, a rapid response satellite. {\it Swift} houses 3 instruments: the Burst Alert Telescope 
(BAT; \cite{bar05}), the X-ray Telescope (XRT; \cite{bur05}) and the Ultra-violet Optical Telescope (UVOT; \cite{roming}). 
The energy ranges of the BAT and the XRT instruments are 15~keV~-~350~keV and 0.2~-~10~keV, respectively, and the wavelength 
range of the UVOT is 1600\AA-8000\AA. The large field of view of the BAT (2 str), enables $\rm 1/6^{th}$ of the sky to 
be searched for GRBs at any one time. Once a GRB has been detected by the BAT, {\it Swift} rapidly slews allowing the XRT and UVOT to 
observe the afterglow within a few tens of seconds after the BAT trigger.

Since launch, {\it Swift} has produced a large sample of UV/optical and X-ray lightcurves which begin soon after the trigger. The high 
detection rate with the XRT \citep[96\%;][]{bur08} allowed a large number of XRT lightcurves to be obtained within the first year. The systematic 
reduction of this sample resulted in the discovery of a 4 segment canonical XRT lightcurve \citep{zhang05, nousek}. After 2 years of operation the 
UVOT, with a much lower detection rate than the XRT  \citep[26\%;][]{roming08}, has detected more than 50 optical afterglows. 
This allows for the first time a systematic reduction and analysis of a significant sample of GRBs with optical afterglows 
observed with the UVOT and allows an investigation of their generic characteristics.

In this paper we present and analyze a sample of 27 UVOT lightcurves of GRB afterglows. In Section 2 we explain how 
we selected the sample of UVOT lightcurves and in Section 3 we describe how we systematically reduced and analyzed 
them. In Section 4 we present the results and in Section 5 we discuss our findings. Throughout the paper we will 
use the following flux convention, $ F\,\propto\,t^{\alpha}\,\nu^{\beta+1}$ with $\alpha$ and $\beta$ being the 
temporal and photon indices respectively. We assume the Hubble parameter $ H_0\,=\,70$\,$\rm km\,s^{-1}\,Mpc^{-1}$ 
and density parameters $\Omega_\Lambda$\,=\,0.7 and $\Omega_m$\,=\,0.3. Unless stated otherwise, all uncertainties 
are quoted at 1$\sigma$. 

\section{The Sample}
To investigate the nature of GRB optical/UV lightcurves a large number of well sampled, good quality UVOT lightcurves 
were required. The sample was selected according to the following specific criteria: the optical/UV lightcurves must 
be observed in the v filter of the UVOT with a magnitude of $\leq$17.8, UVOT observations must have commenced 
within the first 400s after the BAT trigger and the afterglow must have been observed until at least $10^5$s 
after the trigger. These selection criteria ensure the lightcurves have adequate signal to noise and cover 
both early and late time evolution. In addition to the above criteria, the colour of the afterglows must 
not evolve significantly with time, meaning that at no stage should the lightcurve from a single filter significantly deviate 
from any of the other filter lightcurves when normalized to the v filter. This ensures 
that a single lightcurve can be constructed from the UVOT multi-filter observations. Three GRBs, GRB~060218, GRB~060614 and GRB~060729 were 
excluded as they showed significant colour evolution. 

In total 27 GRBs, which occurred between 1st January 2005 and 1st August 2007, fit the selection criteria. As 
there were no short GRBs that met the selection criteria, all the GRBs in this sample are long. Observations, for the majority of 
GRBs in this sample, began within the first 100s and the optical afterglow was detected until at 
least $10^5$s. Formally, GRB~050820a meets our selection criteria, but we have excluded this burst from the 
sample because the BAT triggered on a precursor of this GRB, and the main GRB took place as {\it Swift} 
entered the South Atlantic Anomaly (SAA) with the consequence that UVOT completely missed the early phase of the afterglow.

\section{Data Reduction}
After the BAT has triggered on a GRB and {\it Swift} has slewed, the UVOT performs a sequence of pre-programmed 
exposures of varying length in multiple observing modes and filters designed to balance good time resolution and 
spectral coverage. Observations are performed in either event mode, where the arrival time and position 
are recorded for every photon detected, or in image mode, in which the data are recorded as an image accumulated over 
a fixed period of time. The pre-programmed observations begin with the settling and finding chart exposures. The 
settling exposure is not included in this analysis because the cathode voltage may still increase during the first 
few seconds. Two finding charts follow immediately after the settling exposure and these are observed in event 
mode with the v and white filters (as of the 7th November 2008 the finding charts are observed in u and white). 
The rest of the pre-programmed observations are a combination of event and image mode observations until $\sim$2700s after the 
trigger, after which, only image mode is used. These observations are taken as a series of short, followed by medium and then long 
exposures, which are usually observed with all seven filters of the UVOT (white, v, b, u, uvw1, uvm2, uvw2). 
However, for some targets, the pre-programmed observation sequence may change or not be executed fully because 
of an observing constraint. 

For GRB observations, the earliest part of the lightcurve is expected to show variability over the shortest timescales. As the 
finding charts, which are exposures of 100-400s, contain the earliest observations of the GRB, it is essential 
to obtain lightcurves from these event lists as well as obtaining lightcurves from the images. 

\subsection{Event and Image Data Reduction}
To obtain the best possible lightcurves, we refined the astrometry of the event files before the count rates were obtained. 
The astrometry of the event list was refined by extracting an image every 10s and cross-correlating 
the stars in the image with those found in the USNO-B1 catalogue. The differences in RA and DEC between the stars 
in the image and the catalogue were converted into pixels and then applied to the position of every event in 
the event list during that particular 10s interval. This process was repeated until the end of each event list.  

The images used in this paper were reprocessed by the the Swift Science Data Center (SDC) for the UVOT GRB 
catalogue \citep{roming08}. These images were used because not all the image files in the Swift archive 
have been corrected for modulo-8 fixed pattern noise\footnotemark[1]; only those processed with {\it Swift} 
processing script version 3.9.9 and later have had this correction applied. For a small number of images, 
the aspect correction failed during the SDC processing and in these cases, the images were corrected using 
in-house aspect correction software. 

An aperture of 5$\arcsec$, selected in order to be compatible with the UVOT calibration \citep{poole}, was used to obtain 
the source count rates. However, for sources with a low count rate, it is more precise to use a smaller aperture \citep{poole}. 
Therefore, below a threshold of 0.5 counts per second the source count rates were determined from 3$\arcsec$ radius 
apertures and the count rates were corrected to 5$\arcsec$ using a table of aperture correction factors contained 
within the calibration. The background counts were obtained from circular regions with radii, typically, of 20$\arcsec$. 
These regions were positioned, in each GRB field, over a blank area of sky. For each GRB, the same source and background 
regions were used to determine the count rates from the event lists and the images. The software used can be found in 
the software release, {\sc headas} 6.3.2 and the version 20071106 of the UVOT calibration files.

For each GRB, to maximise the signal to noise of the observed optical afterglow, the lightcurves from each filter in which the 
burst was detected were combined into one overall lightcurve. The lightcurves corresponding to the different filters were 
normalized to that in the v filter. This was possible because there was no significant colour evolution between the filters, 
which was one of the selection criteria described in Section 2. The normalization was determined by fitting a power-law 
to each of the lightcurves in a given time range simultaneously. The power-law indices were constrained to be the same 
for all the filters and the normalizations were allowed to vary between the filters. The ratios of the power-law 
normalizations were then used to normalize the count rates in each filter. The time range used to normalize the 
lightcurves was selected on the basis that the time range included data from all filters in which the burst was detected 
and (as far as possible) the lightcurves were in a power-law decay phase. Once the lightcurves were normalized, they 
were binned by taking the weighted average of the normalized count rates in time bins of $\delta$T/T\,=\,0.2. The 
overall lightcurves were converted from v count rate to v magnitude using the zero point 17.89 \citep{poole}. For 
many GRBs, the trigger time does not represent the true start of the gamma-ray emission. Therefore, the start of 
the gamma-ray emission was chosen as the start time of the T90 parameter. This parameter corresponds to the time in 
which 90\% of the counts in the 15~keV~-~350~keV band arrive at the detector \citep{sak07} and is determined from 
the gamma-ray event data for each GRB, by the BAT processing script. The results of the processing are publicly 
available and are provided for each trigger at http://gcn.gsfc.nasa.gov/swift\_gnd\_ana.html. The difference 
between the trigger time and the start time of the T90 parameter is typically less than a few seconds. However, 
in a minority of cases the difference is much larger, with the largest difference being 133.1s for GRB~050730.

\footnotetext[1]{\tiny http://heasarc.nasa.gov/docs/swift/analysis/UVOT\_swguide\_v2\_2.pdf}

\subsection{Photometric Redshifts}
Spectroscopic redshifts were obtained from the literature for 19 of the GRBs in the sample (see Table \ref{norm}). 
For a further 4 GRBs, it was possible to determine the redshift using an instantaneous UVOT-XRT Spectral Energy Distribution (SED), which was created using 
the method of \cite{schady}. The SEDs were fit with the best fitting model of either a power-law or broken power-law, 
with Galactic and host galaxy absorption and extinction. The Galactic $N_{H}$ was taken from the Leiden/Argentine/Bonn 
(LAB) Survey of Galactic HI \citep{kalberla} and the Galactic extinction was taken from a composite 100$\mu$m map of 
COBE/DIRBE and IRAS/ISSA observations \citep{schlegel}. For the host extinction the Small Magellanic Cloud (SMC) extinction 
curve \citep{pei92} was assumed. The host reddening and absorption, E(B-V), $N_{H}$, and the redshift were left to vary. 
The resulting redshifts can be found in Table \ref{phot}.

\begin{figure}
\begin{center}
\includegraphics[angle=-90,scale=0.35]{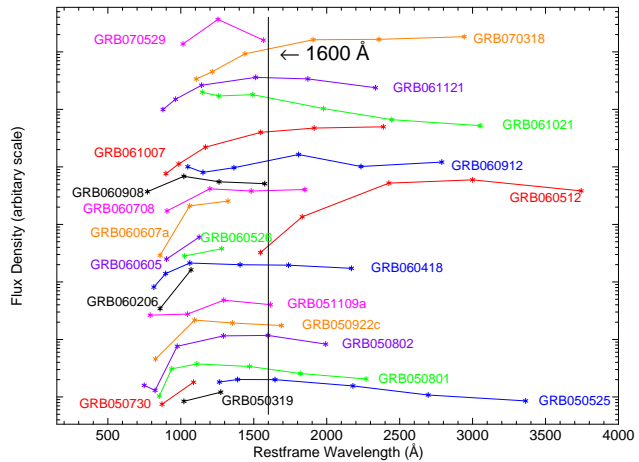}
\caption{The rest frame relative flux SEDs for the 21 GRBs for which a luminosity lightcurve was produced. The relative flux 
for each filter of each GRB was determined by multiplying the relative count rate from Table \ref{norm} by the 
flux conversion factor given in \protect\cite{poole}. The relative flux values have been corrected for Galactic extinction. 
The vertical line at 1600{\AA} marks the rest frame wavelength at which the luminosity lightcurves were produced.}
\label{flux_SED}
\end{center}
\end{figure}

\subsection{Luminosity Lightcurve}
Luminosity lightcurves were produced for all the GRBs whose host E(B-V) value could be determined, except for 
GRBs with photometric redshifts which have a 1$\sigma$ error on the redshift that corresponds to an uncertainty in log luminosity of $>0.1$. 
In total, luminosity lightcurves were produced for 21 of the GRBs in the sample. 

For the 21 GRBs, the observed count rate was converted into luminosity at a common rest frame wavelength. In order to select the common wavelength 
and determine the resulting k-correction factor for each lightcurve, an SED was computed for each GRB. The SED was produced 
by multiplying the relative count rates in each filter, given in Table \ref{norm}, by the count rate-to-flux 
conversion factors given by \cite{poole}. These relative flux densities were corrected for Galactic extinction 
and positioned at the central wavelength of the filter in the rest frame of the GRB; the SEDs are shown in 
Fig. \ref{flux_SED}. The common rest frame wavelength at which to determine the luminosities was selected to 
maximise the number of GRBs with SEDs that include this wavelength and to be relatively unaffected by host extinction. 
The wavelength which best satisfies these conditions is 1600{\AA}. The k-correction factor was taken as the flux density in the rest 
frame at 1600{\AA}, $F_{1600}$, divided by the flux density at the observed central wavelength of 
the v filter (5402{\AA}), $F_{\rm v}$. In the case where a GRB's SED did not cover 1600{\AA}, an average k-correction was 
determined from the other GRBs in the sample, whose SEDs covered both 1600{\AA} and the rest frame wavelength corresponding to the v filter.

To produce the luminosities, the lightcurves in count rate were corrected for Galactic extinction, converted into flux 
density and then into luminosity using the following equation: 
\begin {equation}L(1600)=4\pi D_L^2 F_{\rm v}k\label{luminsoity}\end{equation} 
where $L(1600)$ is the luminosity at a 1600{\AA}, $D_L$ is the luminosity distance, $k=(1+z)(F_{1600}/F_{\rm v})$ 
is the k-correction factor, and $z$ is the redshift of the GRB. Finally, the luminosity lightcurves 
were corrected for host extinction using the $A_{1600}$ values given in Table \ref{norm}. 
These values were determined for the GRBs, using the $A_{\rm v}$ values reported in \cite{schady}.

\begin{figure*}
\begin{center}
\includegraphics[angle=0,scale=0.8]{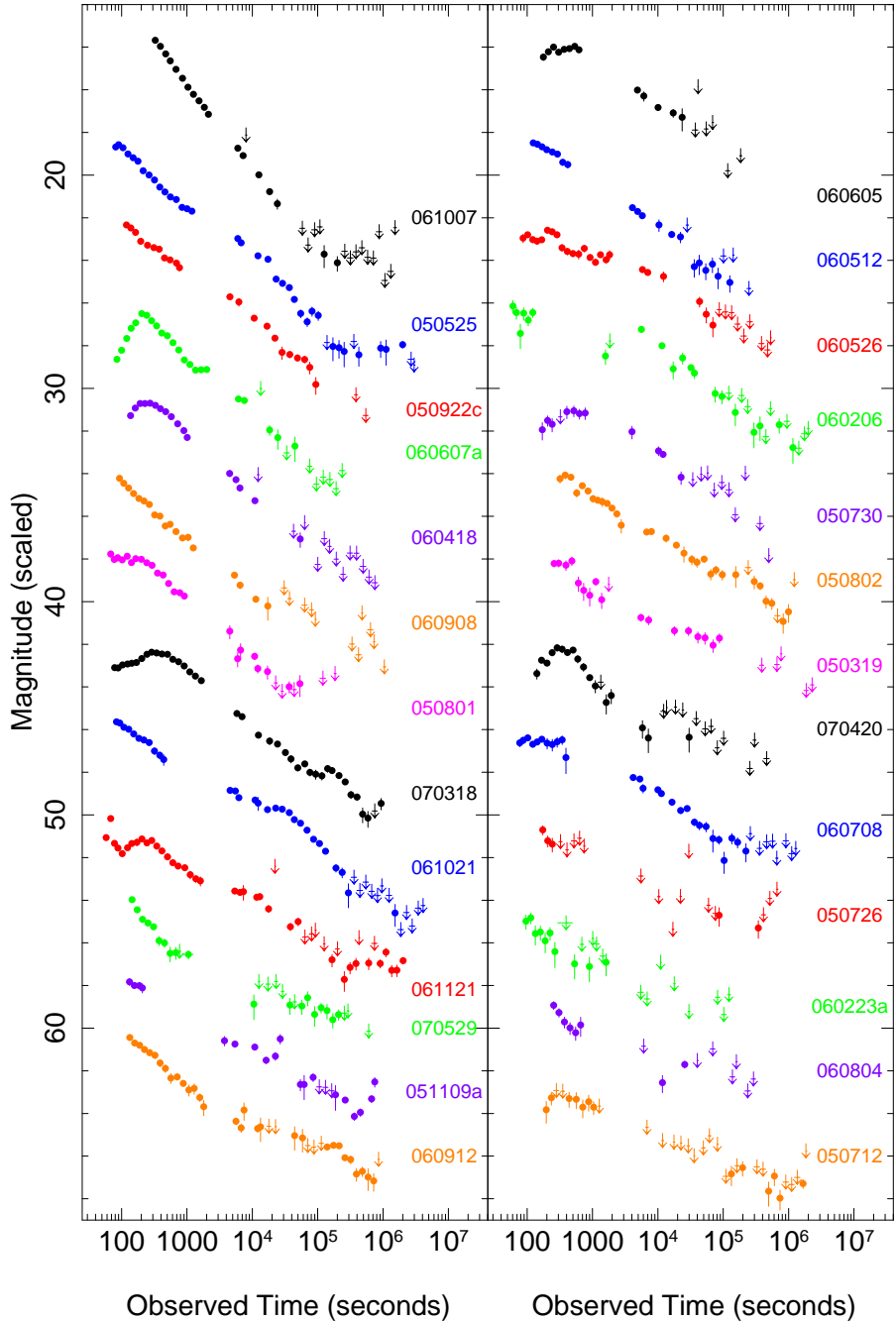}
\caption{The lightcurves, ordered by peak magnitude from brightest (GRB~061007) to faintest (GRB~050712). With a 
few exceptions, the lightcurves appear also to be ordered by decay rate with the brighter bursts decaying more 
rapidly than the fainter bursts. The down arrows in each lightcurve represent 3$\sigma$ upper limits.}
\label{lightcurves}
\end{center}
\end{figure*}
 
\subsection{Bolometric Energy and GRB Classification}

The k-corrected isotropic energy of the prompt gamma-ray emission $E_{k,iso}$, was calculated for each GRB with known redshift 
using Eq. 4 from \cite{bloo01}. The energies were corrected to a rest frame bandpass of 0.1~keV to 10000~keV using one of three 
spectral templates: a power-law, cut-off power-law or Band function \citep{band93}. As Konus-Wind has a larger energy range than BAT, 
the spectral analysis of the prompt emission, observed by Konus-Wind, better represents the spectral behaviour. 
These results were taken from the literature. The spectra observed with Konus-Wind tend to be best fit by a cut-off 
power-law, therefore the spectral template chosen for the GRBs that were not observed with Konus-wind, with power-law spectra of 
photon index $\Gamma>-2$ in the 15~keV~-~150~keV energy range was a cut-off power-law with $E_{peak}$=162.2~keV \citep{D'alessio06}. 
For the GRBs with a power-law spectrum with a photon index of $\Gamma<-2$ in the 15~keV~-~150~keV energy range, a Band function 
was used with $\Gamma_1=-0.99$ \citep{D'alessio06} and $E_{peak}$=15~keV. The resulting k-corrected energies can be found in Table \ref{redshift}. 

The GRBs in this sample were classified into three categories depending on the ratio $R$ of the fluence in the 25~keV~-~50~keV and 
50~-~100~keV BAT energy bands, which are given in \cite{sak07}. The categories and their respective ratios are: an X-ray flash (XRF) 
for $R>1.32$; an X-ray rich GRB (XRR) for $0.72<R\leq1.32$; or a classical GRB (C-GRB) for $R\leq0.72$. Table \ref{fits} lists the GRBs 
with their classifications. In total, there are 12 C-GRBs, 13 XRR and 1 XRF, which is GRB~060512.

\begin{table*}
\begin {tiny}
\begin{tabular}{|@{}l@{}|c|ccccccc}
\hline
&   & & \multicolumn{6}{|c|}{------------Count Rate Relative to v---------------} \\
GRB  & Redshift & $A_{1600}$ & b & u & w1 & m2 & w2 &  white \\
\hline
050319   & $	3.24^{a} $   & 0.27  & 1.36 &  -   &   -  &  -   &  -   & N/A \\
050525   & $	0.606^{b} $  & 0.25  & 2.29 & 2.72 & 1.05 & 0.46 & 0.56 & 4.59\\
050712	 & 	  -	     &  -    &  -   & 0.68 &   -  &  -   &  -   & N/A \\ 
050726   &	  -	     &  -    & 0.63 &  -   &   -  &  -   &  -   & N/A \\ 
050730   & $	3.97^{c} $   & 0.64  & 0.78 &  -   &   -  &  -   &  -   & N/A \\
050801   & $	1.38^{*}$    & 0.00  & 2.25 & 2.46 & 0.81 & 0.29 & 0.13 & N/A \\
050802   & $	1.71^{d} $   & 0.47  & 2.75 & 2.34 & 0.52 & 0.05 & 0.07 & 4.79\\
050922c  & $	2.198^{e} $  & 0.35  & 1.99 & 1.83 & 0.11 &  -   &  -   & N/A \\
051109a  & $	2.346^{f} $  & 0.04  & 1.98 & 0.88 & 0.21 &  -   &  -   & 7.96\\ 
060206   & $	4.04795^{g}$ & 0.00  & 0.42 &  -   &   -  &  -   &  -   & N/A \\
060223a  & $	4.41^{h} $   &  -    & 0.26 &  -   &   -  &  -   &  -   & 4.18\\
060418   & $	1.4901^{i} $ & 0.34  & 1.82 & 1.40 & 0.36 & 0.07 & 0.06 & 6.21\\
060512   & $	0.4428^{j} $ & 1.93  & 3.03 & 2.31 & 0.20 &  -   &  -   & 8.46\\
060526   & $	3.221^{k} $  & 0.00  & 1.39 &  -   &   -  &  -   &  -   & 2.99\\
060605   & $	3.8^{l}  $   & 1.08  & 0.79 &  -   &   -  &  -   &  -   & 2.74\\
060607a  & $	3.082^{m} $  & 0.21  & 1.56 & 0.19 &   -  &  -   &  -   & 4.36\\
060708   & $    1.92^{*}$    & 0.56  & 1.85 & 1.76 & 0.25 &  -   &  -   & 7.49\\
060804   &        -          & -  & 2.10 & 0.00 &   -  &  -   &  -   & N/A\\
060908   & $	2.43^{n} $   & 0.08  & 2.07 & 2.23 & 0.40 &  -   &  -   & 9.27\\
060912   & $	0.937^{o} $  & 2.09  & 1.58 & 2.16 & 0.41 & 0.17 & 0.27 & 8.83\\
061007   & $	1.262^{p} $  & 2.77  & 1.84 & 1.35 & 0.25 & 0.07 & 0.06 & 6.03\\
061021   & $    0.77^{*}$    & 0.00  & 2.38 & 3.14 & 1.74 & 0.81 & 1.23 & 11.97\\
061121   & $    1.314^{q}$   & 1.35  & 2.70 & 2.45 & 0.58 & 0.17 & 0.14 & 9.03\\
070318   & $	0.836^{r} $  & 1.72  & 1.75 & 1.50 & 0.29 & 0.08 & 0.07 & 6.69\\       
070420	 & $	3.01^{*}  $  & 1.53  & 1.52 & 0.47 & 0.19 &  -   &  -   & 6.35\\
070529   & $	2.4996^{s} $ & 0.57  & 3.42 & 0.92 &   -  &  -   &  -   & N/A  \\
\hline							     
\end{tabular}
\end{tiny}
\caption{Spectroscopic redshifts were largely taken from the literature. For four GRBs, photometric 
redshifts, indicated by an *, were determined using the XRT-UVOT SEDs (see Section 3 for more details). The host extinction values at 1600{\AA}, 
were calculated from the best fit $A_{\rm v}$ given in \protect\citep{schady}. The count rate relative to v is provided for each 
filter detected by the UVOT. For a few GRBs, there were no observations with the white filter and for GRB~060804, the white data could not be used; 
the white column for these GRBs contain N/A. References: a) \protect\cite{jak06} b) \protect\cite{3483} c) \protect\cite{3709} d) \protect\cite{3749} e) \protect\cite{jak06} f) \protect\cite{4221} g) \protect\cite{4692} 
h) \protect\cite{4815} i) \protect\cite{5002} j) \protect\cite{5217} k) \protect\cite{jak06} l) \protect\cite{5223} m) \protect\cite{5237} n) \protect\cite{5555} o) \protect\cite{5617} 
p) \protect\cite{5716} q)\protect\cite{5826} r) \protect\cite{6216} s) \protect\cite{6470}}
\label{norm}
\end{table*}   

\section{Results}
In this Section, the flux lightcurves in the observer frame and their properties shall be investigated first, then the luminosity lightcurves 
in the rest frame and their properties shall be examined.

\subsection{Observer Frame Flux Lightcurves}
The lightcurves, shown in Fig. \ref{lightcurves}, are ordered by peak magnitude from the brightest, GRB~061007, 
to the faintest, GRB~050712. Data points with signal to noise below 2 are shown as 3$\sigma$ upper limits. 
The peak magnitude was taken as the maximum magnitude in each binned lightcurve and is given in Table 
\ref{fits}. A trend is observed in the figure, with the brightest GRBs decaying more quickly than the faintest GRBs. 
The lightcurves generally follow one of two types of behaviour. Either they rise to a peak within the first 1000s 
and then decay, or they decay from the beginning of the observations. There are 6 lightcurves that appear to rise to a peak 
between 200s to 1000s after the burst trigger. The peak times for the 6 GRBs were determined from a Gaussian fit to 
each lightcurve in log time. The fit was performed between the brightest data point and the data points on either side 
to the point at which the count rate is 60\% of the peak value. The mean of these peak times is 397s. For the remaining 
GRBs, the beginning of the lightcurve was taken as the upper limit to the peak time and the mean of these upper limits 
is 132s. 

To classify the behaviour of the lightcurves, two power-laws were fit to each lightcurve, covering the time ranges from 
the start of observations until 500s and from 500s until the end of the observations; the results are given in Table 
\ref{fits}. A time of 500s was chosen as it ensures that the early power law fits are performed on at least 100s 
of each lightcurve and because the rising phase tends to occur during the first 500s.

\begin{figure}
\begin{center}
\includegraphics[angle=-90,scale=0.65]{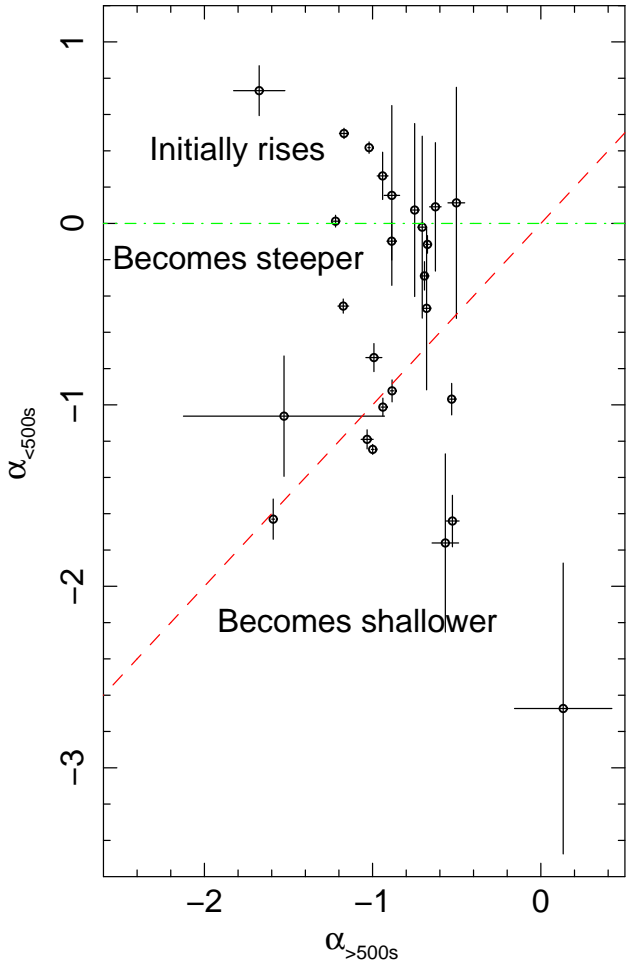}
\end{center}
\caption{Temporal decay after 500s versus temporal decay before 500s. The dashed (red) line indicates the point at which the 
temporal index before 500s equals the temporal index after 500s. The dashed dotted (green) line indicates the cut off between 
GRBs that rise within the first 500s (above the line) and GRBs that decay within the first 500s (below the line).}
\label{versus}
\end{figure}

A comparison of the temporal indices before and after 500s is shown in Fig. \ref{versus}. The figure is divided 
into three groups of behaviour, which are: lightcurves which rise before 500s, lightcurves which decay more 
steeply after 500s and lightcurves which decay less steeply after 500s. Each of these groups contains a similar 
number of GRBs. From Fig. \ref{versus}, there are 4 lightcurves that are clearly rising with an $\alpha_{<500s}$ 
ranging from $0.26 \pm 0.13$ (GRB~060605) to $0.73 \pm 0.14$ (GRB~070420). A further 7 lightcurves are consistent 
with $\alpha_{<500s} \sim 0$ or have large errors and thus it is not clear if these lightcurves are rising, constant 
or decaying before 500s. The remaining 15 lightcurves decay with temporal indices of between $\alpha_{<500s}=-0.12\pm 0.05$ 
(GRB~061121) and $\alpha_{<500s}=-2.67 \pm 0.80$ (GRB~050726). After 500s, all the lightcurves decay with values of 
$\alpha_{>500s}$ ranging from $-0.50\pm0.05$ (GRB~050712) to $-1.67\pm0.15$ (GRB~070420), except for GRB~050726 
where, due to the poor signal to noise, it is not possible to tell if the lightcurve after 500s is rising or decaying.

The mean and intrinsic dispersion of the temporal indices was determined using the maximum likelihood method 
\citep{mac88}, which assumes a Gaussian distribution. The mean for $\alpha_{<500s}$ is $-0.48^{+0.15}_{-0.19}$ with a dispersion of 
$0.69^{+0.19}_{-0.06}$ and the mean for $\alpha_{>500s}$ is $-0.88^{+0.08}_{-0.07}$ with a dispersion 
of $0.31^{+0.07}_{-0.03}$. To see if $\alpha_{<500s}$ and $\alpha_{>500s}$ are independent parameters, a Spearman rank 
test was performed. This test gives a coefficient of -0.22 with a probability of 73$\%$, indicating no evidence for a 
correlation and suggesting that the behaviour after 500s is independent of the behaviour before 500s.

\begin{figure}
\begin{center}
\includegraphics[angle=-90,scale=0.35]{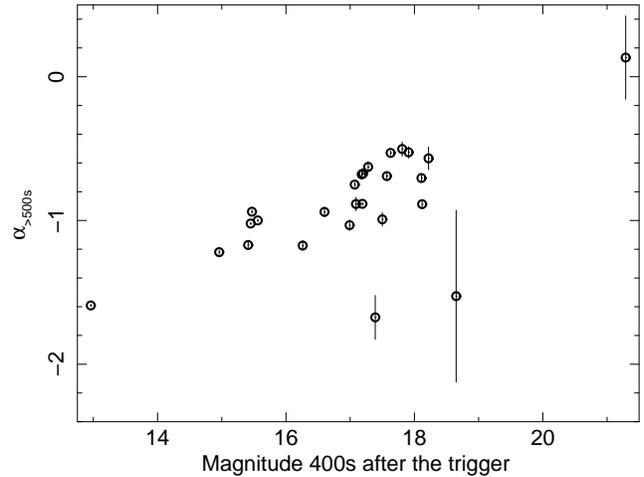}
\caption{Magnitude at 400s against the temporal index after 500s, $\alpha_{>500s}$. A Spearman rank correlation 
test performed on these parameters gives a coefficient of 0.59 at a statistical significance of $99.8\%$, indicating that 
these two parameters are correlated.}
\label{after}
\end{center}
\end{figure}

Since the lightcurves in Fig. \ref{lightcurves} suggest a connection between the brightness and the decay rate, a 
Spearman rank correlation was performed between the temporal indices and the interpolated magnitudes of the 
lightcurves at 400s. The test performed between $\alpha_{<500s}$ and the magnitude at 400s, 
indicates that these parameters are not related, as the coefficient is -0.28 at 84$\%$ confidence. 
However, the Spearman rank test performed on $\alpha_{>500s}$ and the magnitude at 400s gives a coefficient 
of 0.59 at 99.8$\%$ confidence (see Fig. \ref{after}). The correlation is statistically significant at 
$>3\sigma$ and therefore implies that brighter GRBs tend to have faster decays. 

The mean redshifts of the two columns in Fig. \ref{lightcurves} are $\langle{z}\rangle=1.63$ and 
$\langle{z}\rangle=2.98$, suggesting that the decay rate and magnitude are also correlated with redshift because 
the optical afterglows that are brighter and decay more steeply tend to have lower redshifts. A Spearman Rank 
test performed between $\alpha_{>500s}$ and redshift gives no evidence for a correlation with a correlation 
coefficient of 0.07 and a insignificant probability of 23\%. However, a Spearman Rank test performed between the redshift and 
magnitude at 400s, provides a weak correlation with a coefficient of 0.32 and a statistical significance 
of 87\%. Moreover, if the Spearman Rank test is performed between the peak magnitude and redshift, the link 
between the redshift and magnitude is stronger and more significant as the coefficient is 0.55 
and the confidence is 99.3\%. These correlations imply that the correlation between magnitude at 
400s and $\alpha_{>500s}$ is only weakly dependent on redshift.

The lightcurves of the GRBs after 500s, in a few cases, appear to show a change in their temporal behaviour. To quantify 
this behaviour, a broken power-law was fitted to each GRB from 500s until the end of the observations. The broken power-law 
is considered an improvement if the $\chi^2/D.O.F$ has decreased and the probability of chance improvement is small 
($<1\%$), as determined using an F-test. In five cases a broken power-law was an improvement compared 
with a single power-law. The results of the broken power-law fits for the 5 GRBs are given in Table \ref{brkn}. 
In four of these five cases, the broken power-law shows a transition from a shallow to a steeper decay. In the fifth 
case, GRB~070318, the decay became shallower at late times. To test if a single break is sufficient for the decay 
after 500s, a doubly broken power-law was fit to these 5 GRBs. As with the broken power-law, the doubly broken power-law 
was considered an improvement if the $\chi^2/D.O.F$ decreased and the probability of chance improvement is small ($<1\%$). 
The doubly broken power-law was an improvement for only GRB~070318. The best fit values for this model are: 
$\alpha_1=-1.08\pm0.01$, $t_{break,1}=53800^{+6800}_{-6100}$, $\alpha_2=-0.11^{+0.12}_{-0.14}$, 
$t_{break,2}=197000^{+22000}_{-15000}$, $\alpha_3=-1.72\pm0.18$ with $\chi^2/D.O.F=62/23$. 

For the 4 GRBs where the broken power-law was the best fit, the mean decay index before the break is 
$-0.60\pm0.14$ with a dispersion of $0.19^{+0.18}_{-0.07}$ and the mean decay index after the break 
is -1.53$\pm$0.19 with a dispersion of $0.22^{+0.28}_{-0.09}$. The break times range from $6000^{+1000}_{-1100}$s to 
$(4.9\pm0.25)\times 10^{4}$s. If the mean decay index after 500s is recalculated including 
only those lightcurves that decay as a single power-law after 500s, the mean is $-0.87^{+0.10}_{-0.09}$ 
with a dispersion of $0.35^{+0.10}_{-0.04}$. This mean is similar to the mean decay index determined 
using all the GRBs in the sample. For the lightcurves that show a break, the mean decay before the 
break is consistent within 2$\sigma$ with these mean values.

\subsection{GRB Rest Frame Luminosity Lightcurves}

\begin{table}
\begin {tiny}
\begin{center}
\begin{tabular}{|l|c|c}
\hline
GRB  & Photometric Redshift & $\chi^2/D.O.F$\\
\hline
GRB050801   & 1.38  $\pm$ 0.07        & (14/12)   \\     
GRB060708   & 1.92  $\pm$ 0.12        & (21/20)   \\
GRB061021   & $0.77^{+0.06}_{-0.01}$  & (229/174) \\
GRB070420   & $3.01^{+0.96}_{-0.68}$  & (58/60)   \\
\hline
\end{tabular}
\end{center}
\end{tiny}
\caption{Photometric redshifts for 4 of the GRBs without spectroscopic redshifts.}
\label{phot}
\end{table}

\begin{table}
\begin{tiny}
\begin{center}
\begin{tabular}{|l|c|c|@{}c|c|c|c|}
\hline
GRB  &   &$D_L $ (cm) &$\beta_1$ &$\beta_2$ &$E_{peak}$   & $E_{iso,k}$       \\
\hline 				       						   
050319    & BAND & $8.60\times 10^{28}$  & $-0.99$      &   $-2.02^{a}$  & $162.0    $ & 1.44e+53 \\
050525    & CPL  & $1.10\times 10^{28}$  & $-1.10^{b}$  &   $-2.31$        & $ 84.1^{b}$ & 1.59e+53 \\
050730    & CPL  & $1.04\times 10^{29}$  & $-1.53^{a}$  &   $-2.31$        & $162.2    $ & 3.85e+54 \\  
050801    & CPL  & $3.04\times 10^{28}$  & $-1.99^{a}$  &   $-2.31$        & $162.2    $ & 7.67e+53 \\
050802    & CPL  & $3.96\times 10^{28}$  & $-1.54^{a}$  &   $-2.31$        & $162.2    $ & 5.86e+53 \\
050922c   & CPL  & $5.38\times 10^{28}$  & $-1.55^{c}$  &   $-2.31$        & $162.2    $ & 4.30e+54 \\
051109a   & CPL  & $5.82\times 10^{28}$  & $-1.25^{d}$  &   $-2.31$        & $161.0^{d}$ & 6.43e+53 \\
060206    & CPL  & $1.12\times 10^{29}$  & $-1.20^{a}$  &   $-2.31$        & $ 78.0^{a}$ & 3.42e+53 \\
060223a   & CPL  & $1.24\times 10^{29}$  & $-1.74^{a}$  &   $-2.31$        & $162.2    $ & 4.47e+54 \\
060418    & CPL  & $3.34\times 10^{28}$  & $-1.50^{e}$  &   $-2.31$        & $230.0^{e}$ & 2.94e+54 \\  
060512    & BAND & $7.56\times 10^{27}$  & $-0.99$      &   $-2.48^{a}$  & $162.2    $ & 5.04e+50 \\
060526    & BAND & $8.54\times 10^{28}$  & $-0.99$      &   $-2.01^{a}$  & $162.2    $ & 1.37e+53 \\
060605    & CPL  & $1.04\times 10^{29}$  & $-1.55^{a}$  &   $-2.31$        & $162.2    $ & 1.15e+54 \\  
060607a   & CPL  & $8.10\times 10^{28}$  & $-1.47^{a}$  &   $-2.31$        & $162.2    $ & 1.81e+54 \\  
060708    & CPL  & $4.56\times 10^{28}$  & $-1.68^{a}$  &   $-2.31$        & $162.2    $ & 3.79e+53 \\	     
060908    & CPL  & $6.08\times 10^{28}$  & $-1.00^{a}$  &   $-2.31$        & $151.0^{a}$  & 1.88e+53 \\
060912    & POWER& $1.88\times 10^{28}$  & $-1.74^{f}$  &   $-2.31$        & $162.2    $ & 2.52e+54 \\
061007    & BAND & $2.72\times 10^{28}$  & $-0.70^{g}$  &   $-2.61^{g}$    & $399.0^{g}$  & 2.00e+53 \\
061021    & CPL  & $1.48\times 10^{28}$  & $-1.22^{h}$  &   $-2.31$        & $777.0^{h}$ & 9.65e+52 \\	 
061121    & CPL  & $2.86\times 10^{28}$  & $-1.32^{i}$  &   $-2.31$        & $606.0^{i}$ & 2.75e+54 \\  
070318    & CPL  & $1.63\times 10^{28}$  & $-1.42^{a}$  &   $-2.31$        & $162.2    $ & 8.78e+52 \\  
070420    & CPL  & $7.87\times 10^{28}$  & $-1.23^{j}$  &   $-2.31$        & $147.0^{j}$ & 6.39e+54 \\  
070529    & CPL  & $6.14\times 10^{28}$  & $-1.34^{a}$  &   $-2.31$        & $162.2    $ & 6.46e+53 \\
\hline
\end{tabular}
\end{center}
\end{tiny}
\caption{Properties of the GRBs with spectroscopic or photometric redshifts. This table contains the luminosity distance, 
the gamma-ray photon indices and peak energies used to determine the k-corrected isotopic energy for each GRB in the energy range 10~keV~-~10~MeV. 
References: a)\protect\cite{sak07}, b)\protect\cite{3474}, c)\protect\cite{4030} d)\protect\cite{4238}, e)\protect\cite{4989}, f)\protect\cite{5570}, g)\protect\cite{5722}, h)\protect\cite{5748}, i)\protect\cite{5837}, j)\protect\cite{6344} }
\label{redshift} 
\end{table}

\begin{figure*}
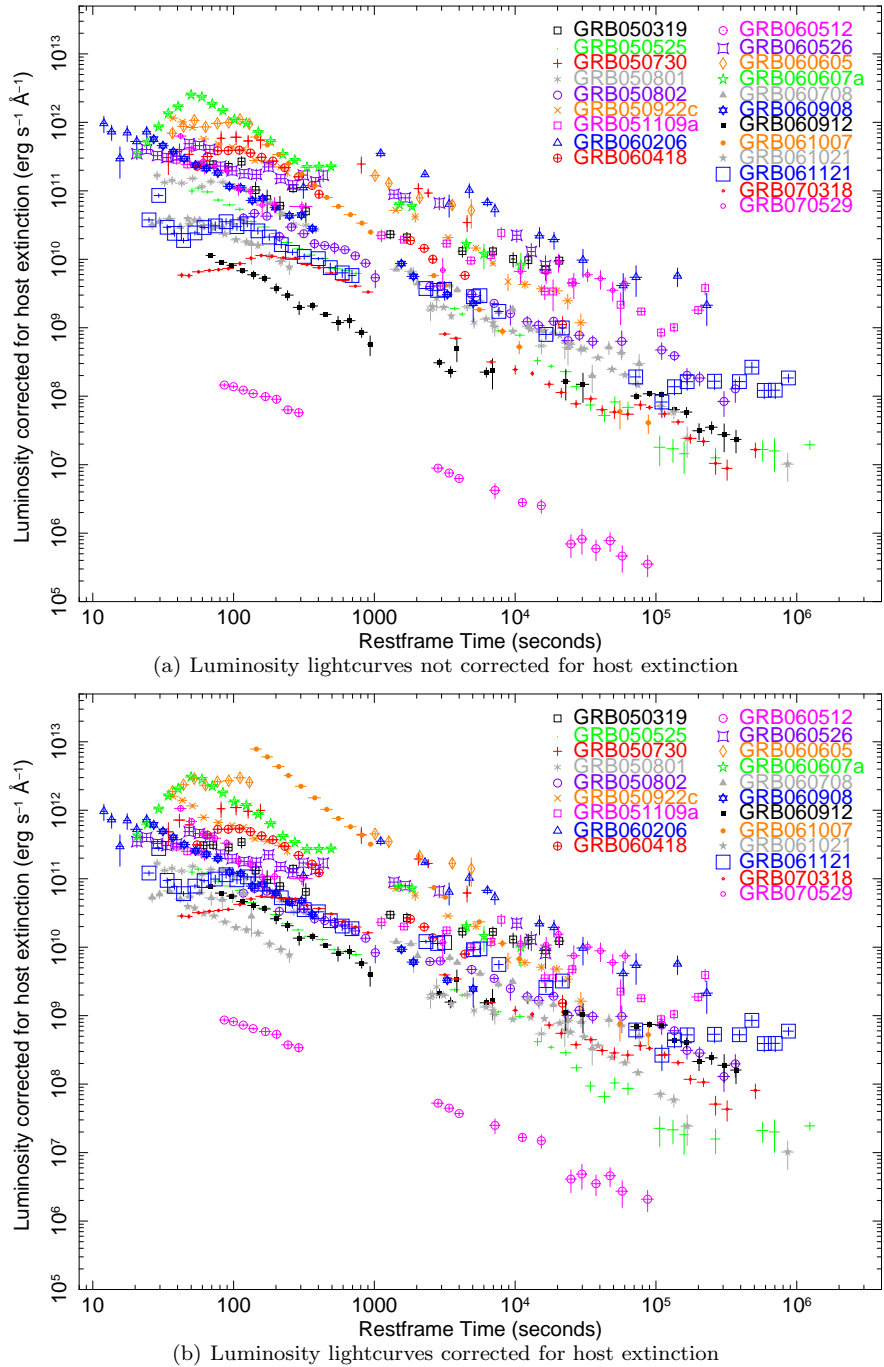

\begin{center}
\subfigure[Luminosity lightcurves not corrected for host extinction]{\label{fig:a}\includegraphics[angle=-90,scale=0.5]{./Images/luminosity_GRBs.cps}}
\subfigure[Luminosity lightcurves corrected for host extinction]{\label{fig:b}\includegraphics[angle=-90, scale=0.5]{./Images/luminosity_host_GRBs.cps}} \\
\end{center}
\caption{The luminosity lightcurves for the 21 lightcurves in the sample with spectroscopic or photometric redshifts. 
Panel A) shows the luminosity lightcurves before correction for the host extinction whereas Panel B) shows the 
luminosity lightcurves after correction for the host extinction. The grey lightcurves are the lightcurves with photometric redshifts.}
\label{luminosity}
\end{figure*}

The luminosity lightcurves at 1600{\AA}, in units of erg s$^{-1}$ cm$^{-1}$ $\rm \AA^{-1}$, before and after correction 
for host extinction, are shown in Fig. \ref{luminosity}.
Panel (a) shows the luminosity lightcurves before any correction for the host extinction has been applied. In both panels 
of Fig. \ref{luminosity}, GRB~060512 lies significantly below all the other lightcurves. We suspect that this is caused 
by either an incorrect determination of the host extinction or of the redshift. The redshift of this GRB could be 
wrong because it was not determined from the afterglow spectra, but was based on the alignment of the GRB 
with a galaxy \citep{5217}. In \cite{schady}, the best fitting model to the SED of GRB~060512 gives a poor fit 
with a $\chi^2/D.O.F=84/23$ and a host extinction of $E(B-V)_{host}=0.16^{+0.01}_{-0.00}$. A photometric redshift 
was determined for this GRB using the method described in Section 2, of $z=2.279^{+0.09}_{-0.18}$ and an 
extinction of $A_{1600}=0.00^{+0.02}_{-0.00}$. Using these values a luminosity lightcurve for GRB~060512 at 
1600{\AA} was produced. This photometric redshift changes the rest frame relative flux SED and consequently the k-correction factor. 
The result is that the luminosity lightcurve increases by $\sim3$ orders of magnitude, which means that this GRB is no longer 
separated from the rest of the GRBs in the sample. Nonetheless, this GRB may be intrinsically different to 
all the other GRBs in the sample as this GRB is the only XRF in the sample and it may be that XRFs are a 
class of sub-luminous GRBs. However, as it is uncertain whether the redshift of GRB~060512 is correct, the luminosity 
lightcurve for this GRB will be excluded from any further analysis.

\begin{figure*}
\subfigure[]{\label{fig:edge-d}\includegraphics[scale=0.32]{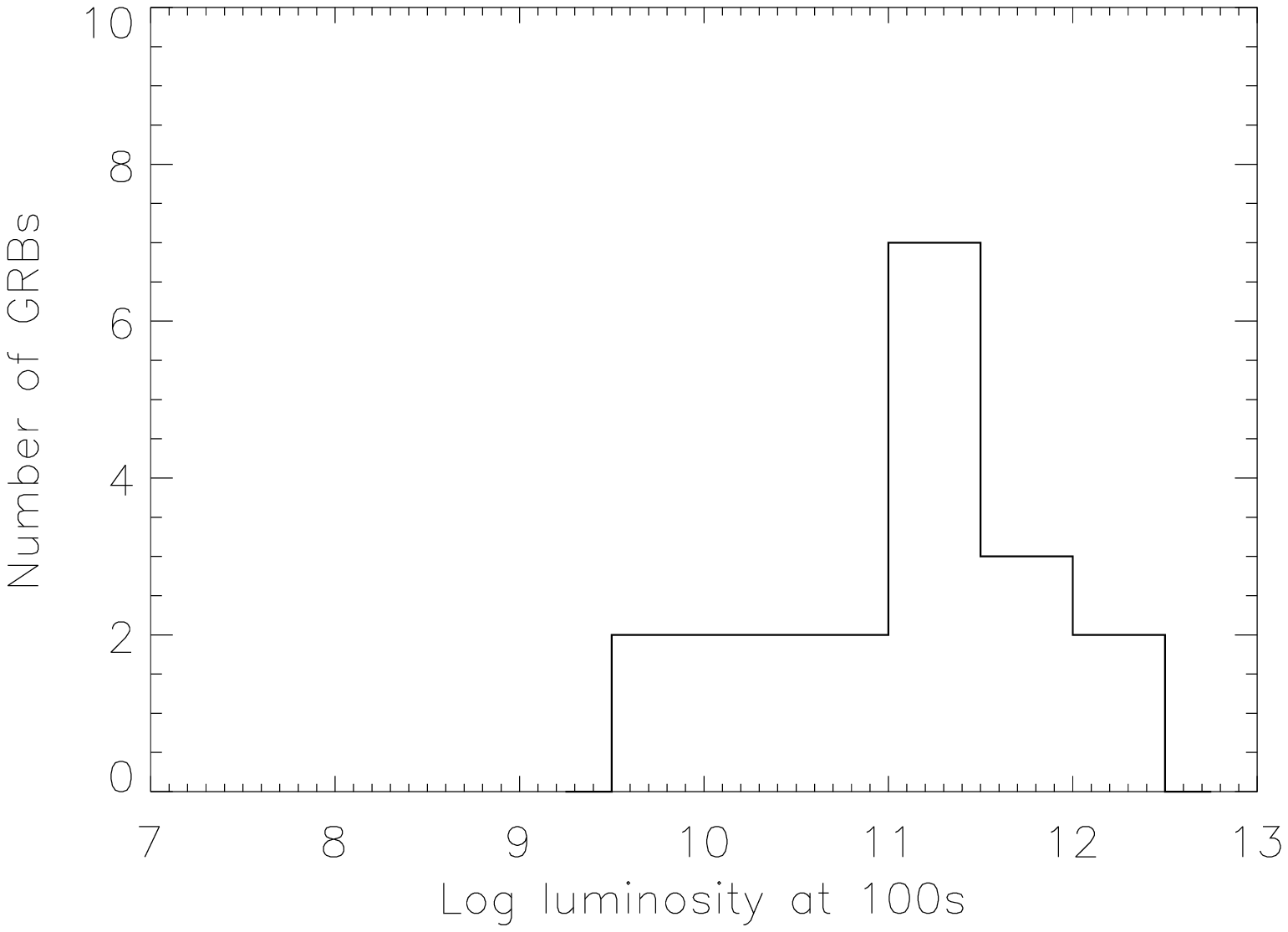}}
\subfigure[]{\label{fig:edge-e}\includegraphics[scale=0.32]{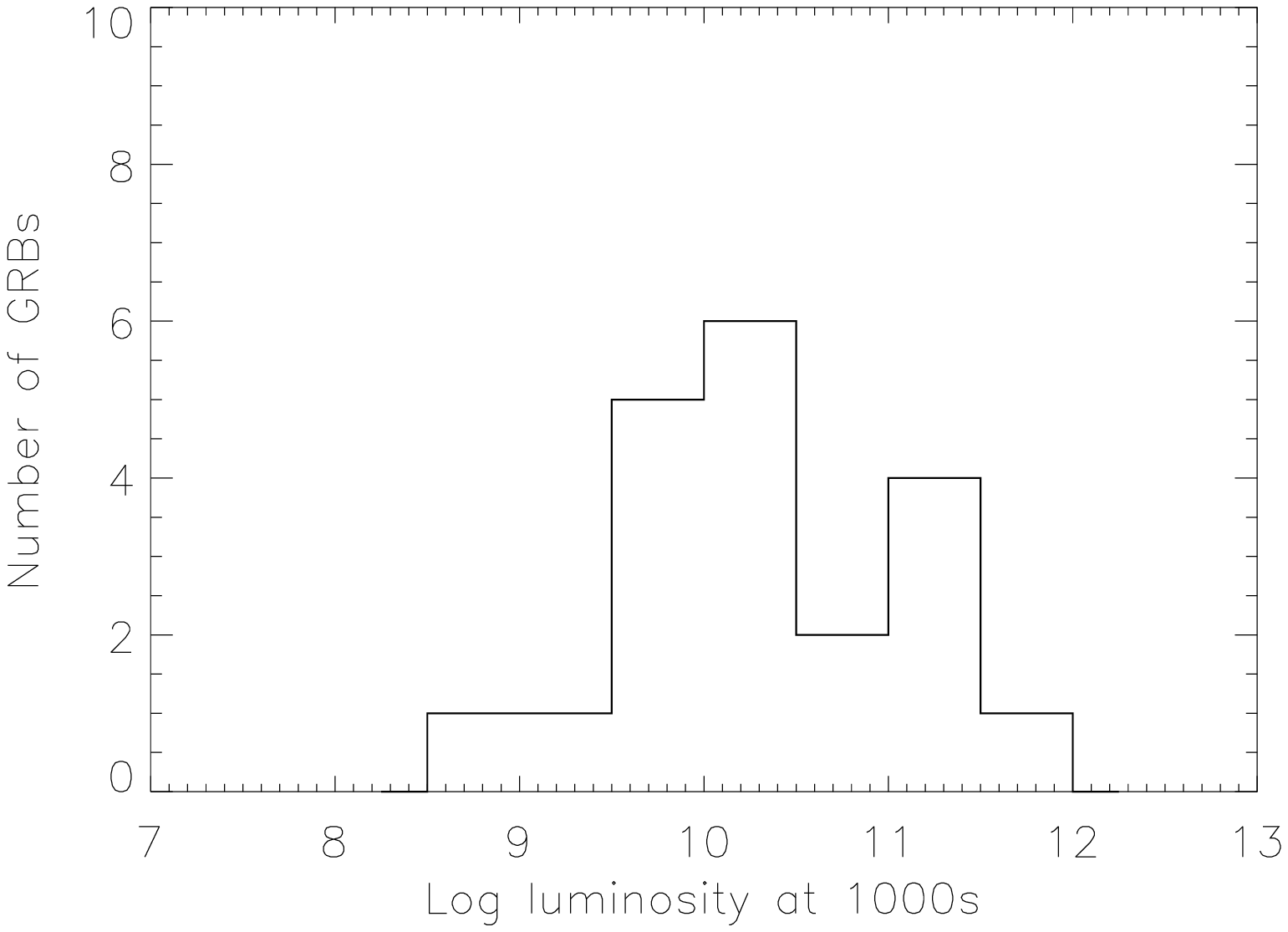}} 
\subfigure[]{\label{fig:edge-f}\includegraphics[scale=0.32]{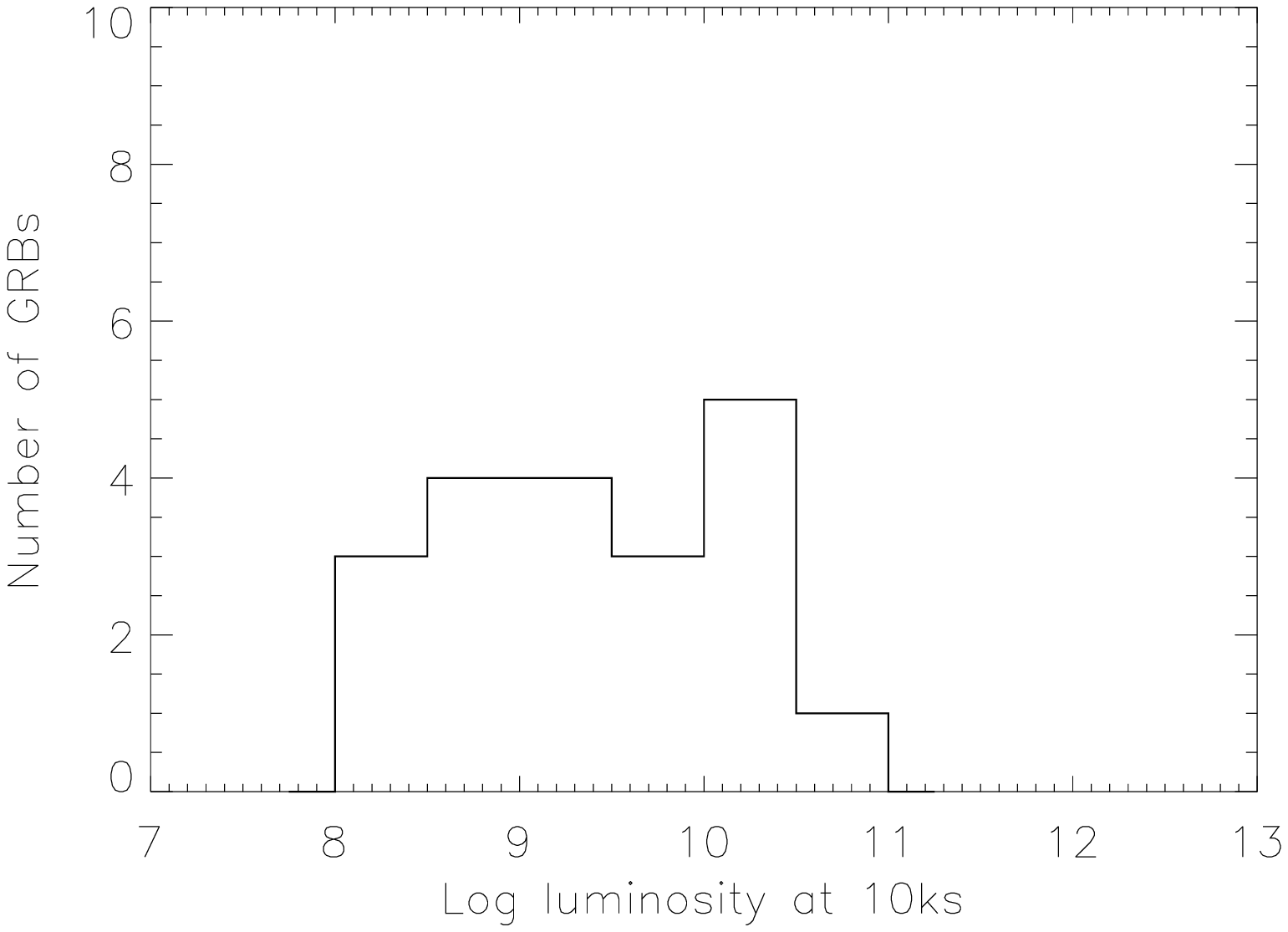}} \\
\subfigure[]{\label{fig:edge-a}\includegraphics[scale=0.32]{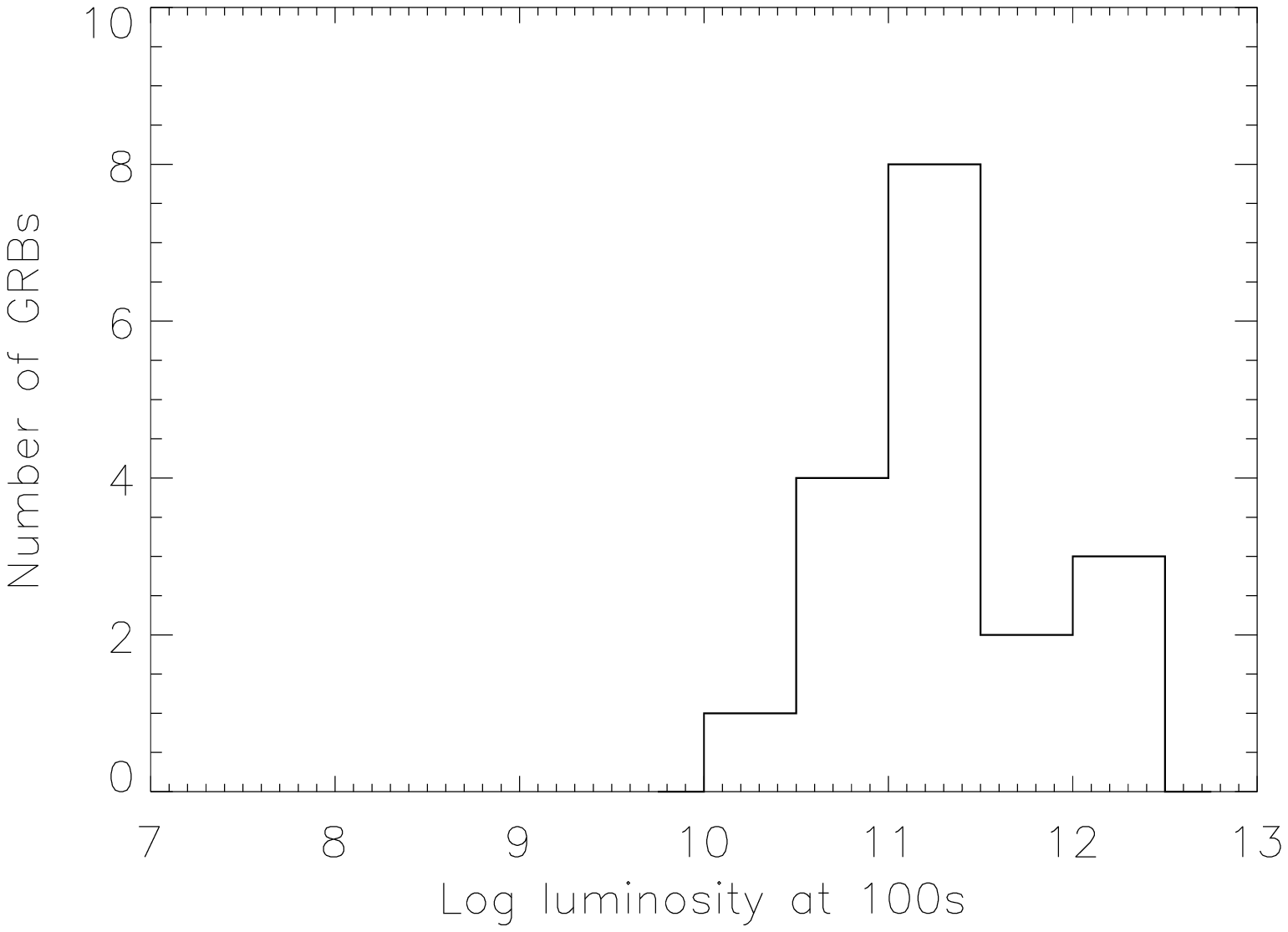}}
\subfigure[]{\label{fig:edge-b}\includegraphics[scale=0.32]{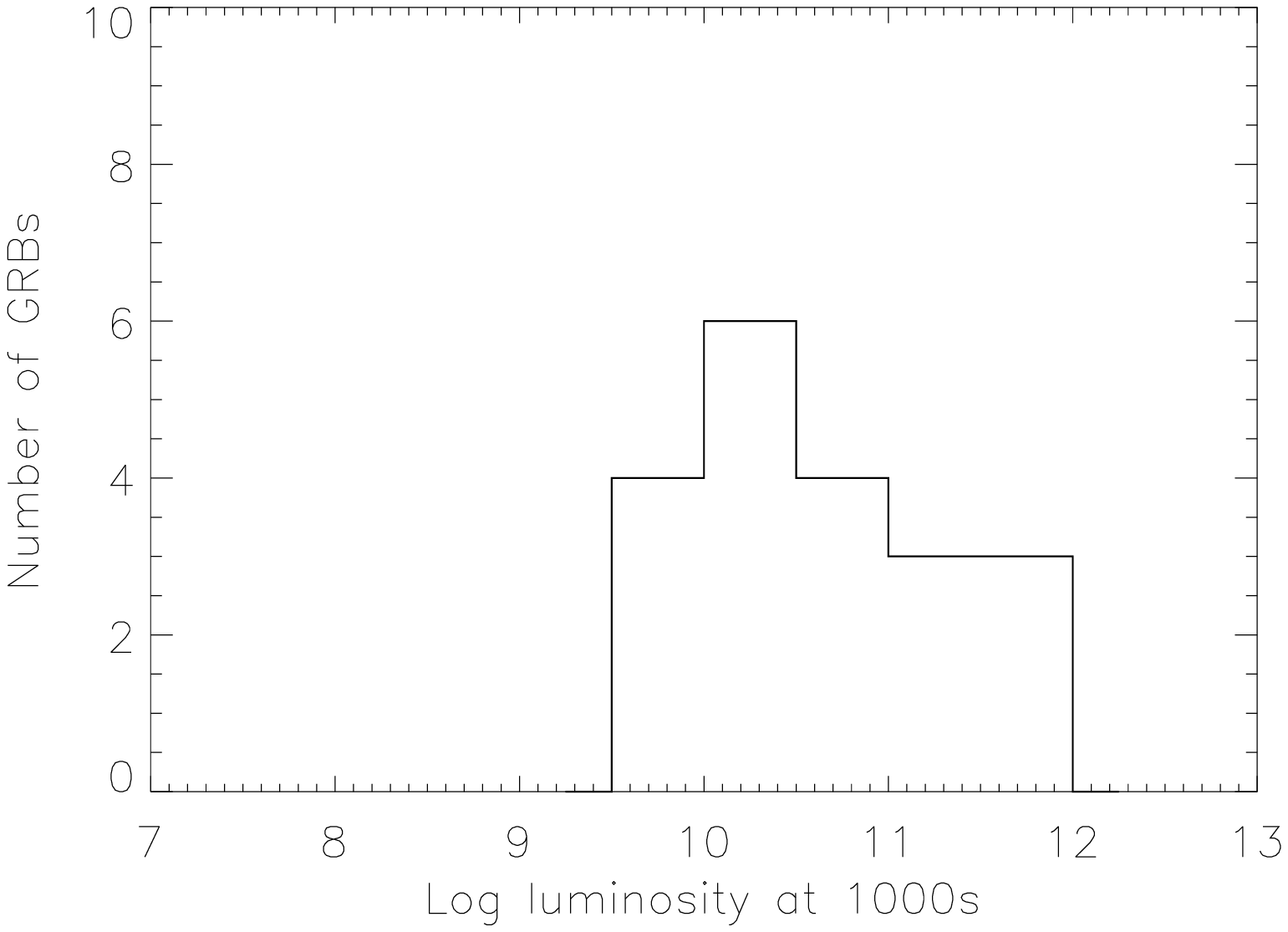}} 
\subfigure[]{\label{fig:edge-c}\includegraphics[scale=0.32]{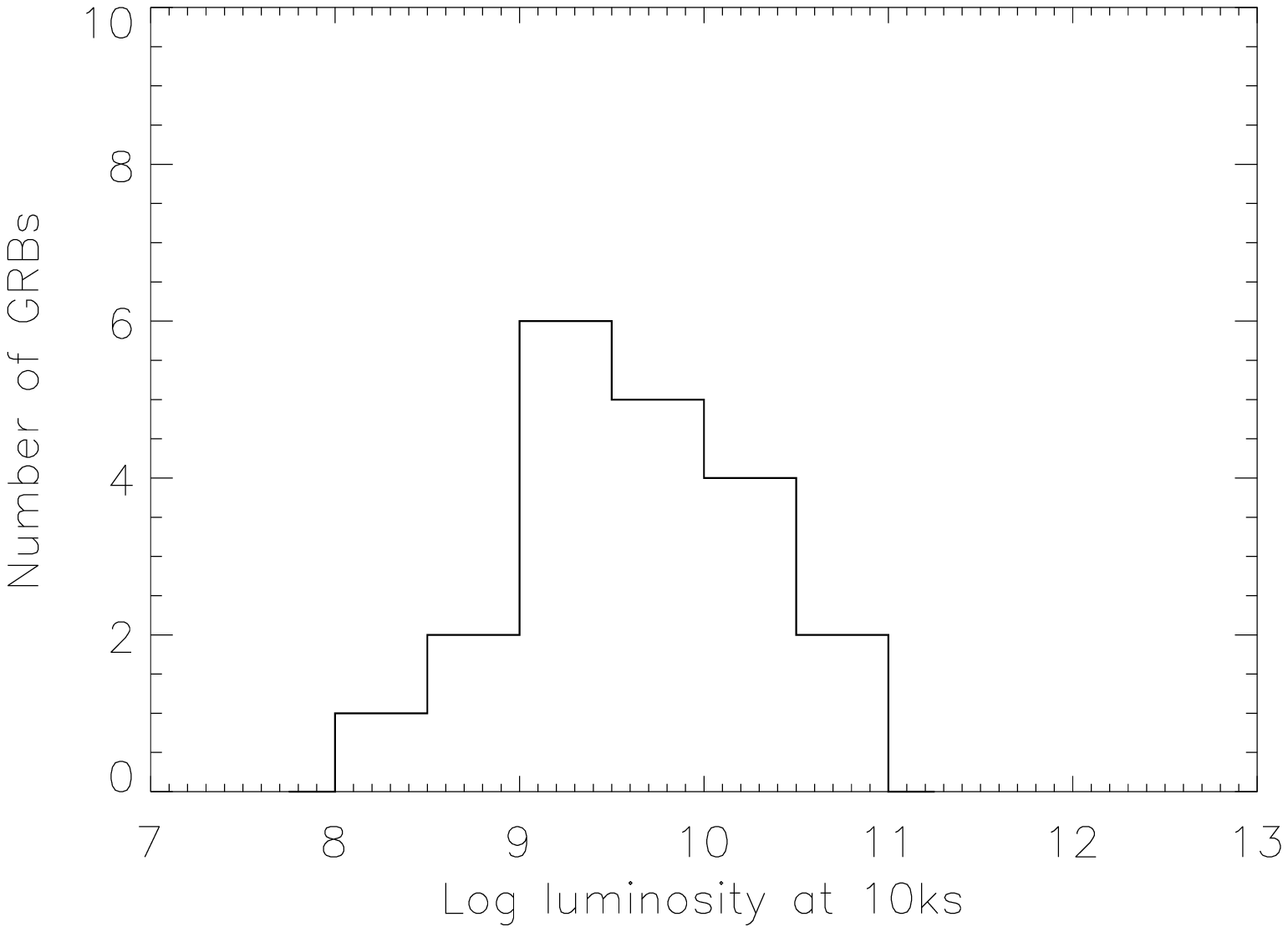}} \\
\caption{Luminosity Distributions. The luminosities in panels (a) to (c) show the luminosities at rest frame 100s, 1000s and 10~ks 
that have not been corrected for host extinction. Panels (d) to (f) show the luminosity distributions 
at the same epochs, but with the lightcurves corrected for host extinction.}
\label{lum_hist}
\end{figure*}

To produce luminosity distributions, the luminosities were interpolated from the lightcurves 
before and after correction for host extinction at the 3 rest frame epochs: 100s, 1000s and 10~ks. The logarithmic distribution of the 
luminosities at the three epochs are shown in Panels (a) to (f) of Fig. \ref{lum_hist}. The distributions consisting 
of the luminosities at 100s contain 18 GRBs whereas the distributions for the luminosities at 1000s and 10~ks contain 20 GRBs. Panels 
(a) to (c), show the logarithmic distribution of luminosities before correction for host extinction. The means of these 
distributions at 100s, 1000s and 10~ks in the rest frame are 11.08, 10.29 and 9.39, respectively. The standard deviations at the three rest frame 
epochs are 0.65 at 100s, 0.71 at 1000s, and 0.68 at 10~ks. Panels (d) to (f) show the logarithmic distributions 
of the luminosity lightcurves that have been corrected for host extinction. The mean of the 
host extinction corrected distributions at 100s, 1000s and 10~ks in the rest frame are 11.29, 10.55 and 9.64, 
respectively and the standard deviations are 0.57 at 100s, 0.67 at 1000s, and 0.62 at 10~ks.

\begin{table*}
\begin {tiny}
\begin{tabular}{|l|cc@{}|r|cc|c|c|}
\hline
Name  & Classification  &Peak Magnitude & Peak Time (s) &$\alpha_{<500s}$ & $(\chi^2/D.O.F)$ & $\alpha_{>500s}$ & ($\chi^2/D.O.F$)\\
\hline
050319  &  XRR     & 17.09 & < 234  &  0.09 $\pm$ 0.35 &   0/ 2 & -0.63 $\pm$ 0.03 &  34/17	   \\	 
050525  &  XRR     & 13.57 & < 78   & -1.25 $\pm$ 0.03 &  21/ 9 & -1.00 $\pm$ 0.01 & 398/26   \\ 
050712  &  C-GRB   & 17.77 & < 178  &  0.11 $\pm$ 0.64 &   4/ 3 & -0.50 $\pm$ 0.05 &  52/26	   \\ 
050726  &  C-GRB   & 17.21 & < 159  & -2.67 $\pm$ 0.80 &   1/ 3 &  0.13 $\pm$ 0.29 &  21/13	   \\
050730  &  C-GRB   & 17.22 &   744  &  0.15 $\pm$ 0.50 &   6/ 3 & -0.89 $\pm$ 0.05 &  72/16	   \\
050801  &  XRR     & 15.26 & < 66   & -0.46 $\pm$ 0.04 &  43/10 & -1.17 $\pm$ 0.03 &  41/15	   \\
050802  &  XRR     & 17.07 & < 289  &  0.07 $\pm$ 0.48 &   0/ 1 & -0.75 $\pm$ 0.01 &  53/29   \\   
050922c &  C-GRB   & 14.34 & < 109  & -1.01 $\pm$ 0.05 &  21/ 6 & -0.94 $\pm$ 0.01 &  88/14	   \\
051109a &  C-GRB   & 16.33 & < 122  & -0.47 $\pm$ 0.45 &   0/ 2 & -0.68 $\pm$ 0.02 &  80/16	   \\
060206  &  XRR     & 16.64 & < 57   & -0.02 $\pm$ 0.50 &   9/ 4 & -0.70 $\pm$ 0.03 & 205/22   \\
060223a &  XRR     & 17.33 & < 88   & -1.06 $\pm$ 0.33 &   4/ 6 & -1.53 $\pm$ 0.60 &  31/14	   \\   
060418  &  XRR     & 14.69 &   260  &  0.01 $\pm$ 0.03 & 138/ 6 & -1.22 $\pm$ 0.01 &  58/21	   \\
060512  &  XRF     & 16.50 & < 114  & -0.74 $\pm$ 0.08 &   3/ 6 & -0.99 $\pm$ 0.05 &  10/14	   \\
060526  &  XRR     & 16.59 & < 82   & -0.29 $\pm$ 0.08 &  36/ 8 & -0.69 $\pm$ 0.02 & 141/22   \\
060605  &  XRR     & 16.50 &   459  &  0.26 $\pm$ 0.13 &   6/ 4 & -0.94 $\pm$ 0.03 &  27/11	   \\
060607a &  C-GRB   & 14.50 &   254  &  0.50 $\pm$ 0.03 & 702/ 8 & -1.17 $\pm$ 0.03 & 152/20   \\
060708  &  XRR     & 17.19 & < 72   & -0.10 $\pm$ 0.10 &   8/ 9 & -0.89 $\pm$ 0.03 &  39/23	   \\
060804  &  XRR     & 17.43 & < 231  & -1.76 $\pm$ 0.49 &   0/2  & -0.57 $\pm$ 0.08 &  21/9  \\
060908  &  C-GRB   & 15.21 & < 88   & -1.19 $\pm$ 0.05 &   6/ 8 & -1.03 $\pm$ 0.04 &  20/18	   \\   
060912  &  XRR     & 16.44 & < 114  & -0.97 $\pm$ 0.09 &   2/ 6 & -0.53 $\pm$ 0.01 & 149/27   \\
061007  &  C-GRB   & 12.68 & < 298  & -1.63 $\pm$ 0.11 &   1/ 1 & -1.59 $\pm$ 0.01 & 129/28   \\   
061021  &  C-GRB   & 15.64 & < 79   & -0.92 $\pm$ 0.06 &   3/ 9 & -0.88 $\pm$ 0.01 & 556/29   \\
061121  &  C-GRB   & 15.67 & < 53   & -0.12 $\pm$ 0.05 & 181/11 & -0.67 $\pm$ 0.01 & 179/31   \\
070318  &  C-GRB   & 15.37 &   316  &  0.42 $\pm$ 0.03 &  19/ 9 & -1.02 $\pm$ 0.01 & 261/27   \\
070420  &  C-GRB   & 17.16 &   347  &  0.73 $\pm$ 0.14 &  13/ 5 & -1.67 $\pm$ 0.15 &  56/23	   \\
070529  &  C-GRB   & 15.95 & < 131  & -1.64 $\pm$ 0.14 &   5/ 5 & -0.53 $\pm$ 0.04 &  42/19   \\
\hline
\end{tabular}
\end{tiny}
\caption{The v peak magnitude, peak time and best fit decay indices of the lightcurves before and after 500s and the classification of each GRB. 
In the cases where a peak was not observed an upper limit to the peak time is provided. The GRBs are classified 
into three categories depending on the ratio of the fleuence in the 25~keV~-~50~keV and 50~-~100~keV BAT energy 
bands, which are given in \protect\cite{sak07}: an X-ray flash (XRF), an X-ray rich GRB (XRR) or a classical GRB (C-GRB).}
\label{fits}
\end{table*}

\begin{figure}
\includegraphics[angle=0,scale=0.55]{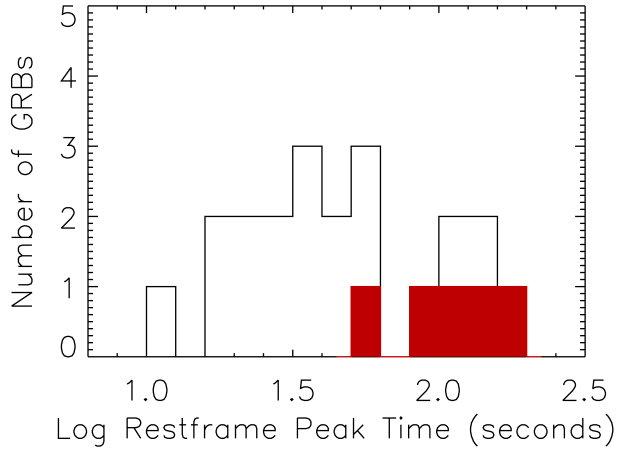}
\caption{Distribution of the rest frame peak times. The red filled area represents the GRBs with known rest frame peak times, 
whereas the unfilled area contains the GRBs with known rest frame peak times and those GRBs with only upper limits to their peak time. 
Only 21 GRBs, for which luminosity lightcurves were produced, are included in this figure.}
\label{time_hist}
\end{figure}

The distribution of rest frame peak times and upper limits is shown in Fig. \ref{time_hist}. The peak 
times of the GRBs with observed peaks overlap with the upper limits of the GRBs without observed peaks. 
Therefore, it is not possible to tell if the GRBs with observed peaks are a separate class, or if they 
belong to the tail end of a distribution where the majority of GRB peaks occur before the UVOT can observe them.

\begin{figure}
\includegraphics[angle=-90,scale=0.32]{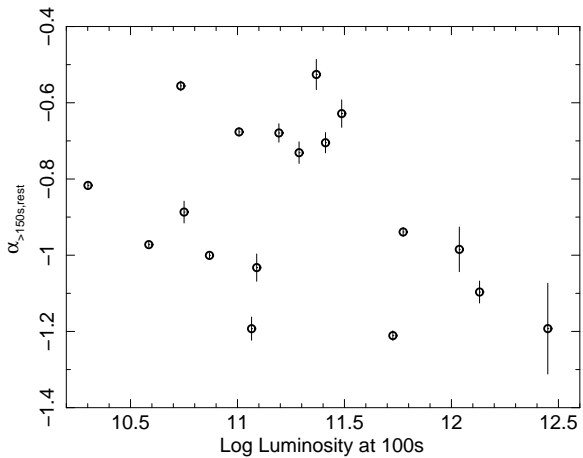}
\caption{Temporal index determined from the lightcurves in the rest frame from 150s onwards, shown against luminosity of the lightcurves at 100s.}
\label{after_150s}
\end{figure}

To determine if the relationship between the brightness of the afterglow and the late time decay rate is intrinsic, the 
luminosity lightcurves were fitted with a power-law from 150s onwards $\alpha_{>150s,rest}$, where $150{\rm s}\simeq 500{\rm s}/(1+\langle z\rangle)$ 
and $\langle z\rangle=2.21$ is the mean redshift of the GRBs in the sample, and a Spearman rank test was performed between this 
decay and the extinction corrected luminosity at 100s in the rest frame. These parameters are shown in Fig. \ref{after_150s}. The Spearman rank 
test does not support or refute a correlation between these parameters because the coefficient is -0.29 and the probability of 
correlation is not significant at 76\%. 

\begin{table*}
\begin {tiny}
\begin{tabular}{|l|c|c|c|c|c}
\hline
GRB  & $\alpha_{1}$ & Break Time & $\alpha_{2}$ & ($\chi^2/D.O.F$) & $\Delta\chi^2$\\
\hline
GRB~050525 &  $-0.80 \pm 0.01$  &  $16400  ^{+1200 }_{-1400}$ & $-1.70 \pm 0.08$        &  ( 120/24 ) & 278 \\
GRB~050922c&  $-0.76 \pm 0.03$  &  $6000   ^{+1100}_{-1000}$  & $-1.20 \pm 0.05 $       &  ( 16/12 )  & 59  \\
GRB~060526 &  $-0.33 \pm 0.04$  &  $30800  ^{+4700}_{-5800}$  & $-2.33^{+0.65}_{-0.47}$ &  ( 32/20 )  & 93  \\
GRB~061021 &  $-0.51 \pm 0.02$  &  $49300  ^{+2500}_{-2500}$  & $-1.60^{+0.07}_{-0.06}$ &  ( 49/27 )  & 507 \\
GRB~070318 &  $-1.09 \pm 0.01$  &  $16000  ^{+3600}_{-3000}$  & $-0.78 \pm 0.03       $ &  ( 131/25 ) & 129 \\
\hline
\end{tabular}
\end{tiny}
\caption{Five of the lightcurves in the sample could be fit with a broken power-law from 500s onwards. The final 
column provides the difference in $\chi^2$ (for 2 additional degrees of freedom) when fitting the lightcurve 
after 500s with a single and broken power laws.}
\label{brkn}
\end{table*}

\begin{table}
\begin{tiny}
\begin{tabular}{|l|c|c|c|}
\hline
GRB   & Lorentz Factor at Peak & $M_{fb}$ & $R_{dec}$ \\
\hline 				       						   
050319    & >168 & <2.40e-04  &  <9.33e+16  \\
050525    & >179 & <2.49e-04  &  <9.25e+16  \\
050730    &  174 &  6.18e-03  &   2.72e+17  \\  
050801    & >275 & <7.81e-04  &  <1.17e+17  \\
050802    & >156 & <1.05e-03  &  <1.56e+17  \\
050922c   & >308 & <3.91e-03  &  <1.93e+17  \\
051109a   & >236 & <7.61e-04  &  <1.22e+17  \\
060206    & >338 & <2.83e-04  &  <7.80e+16  \\
060223a   & >409 & <3.06e-03  &  <1.62e+17  \\
060418    &  193 &  4.27e-03  &   2.32e+17  \\  
060512    & > 72 & <1.95e-06  &  <2.48e+16  \\
060526    & >247 & <1.55e-04  &  <7.09e+16  \\
060605    &  177 &  1.82e-03  &   1.80e+17  \\  
060607a   &  220 &  2.30e-03  &   1.81e+17  \\  
060708    & >256 & <4.14e-04  &  <9.72e+16  \\	     
060908    & >231 & <2.27e-04  &  <8.23e+16  \\
060912    & >235 & <3.00e-03  &  <1.94e+17  \\
061007    & >126 & <4.42e-04  &  <1.26e+17  \\
061021    & >173 & <1.56e-04  &  <7.99e+16  \\	 
061121    & >337 & <2.28e-03  &  <1.57e+17  \\   
070318    &  103 &  2.38e-04  &   1.09e+17  \\  
070420    &  228 &  7.84e-03  &   2.69e+17  \\  
070529    & >234 & <7.71e-04  &  <1.23e+17  \\
\hline
\end{tabular}
\end{tiny}
\caption{Properties of the GRBs, derived assuming the rise of the forward shock is the cause of the rise observed in the UVOT lightcurves. 
The initial Lorentz factors were determined using the peak times and Eq. 1 of \protect\cite{mol06}. Where only an upper limit to the peak time 
is known, only a lower limit to the Lorentz factor is given. The fraction of mass as baryons and the deceleration radius are 
determined using the Lorentz factor.} 
\label{energies}
\end{table}

\section{Discussion}
In this section, we shall discuss the possible mechanisms that could produce the 
rising behaviour of the early afterglow viewed in the observer frame before 500s, and we shall discuss the late 
afterglow from 500s onwards. We will also discuss the implications of the luminosity distribution 
and compare the UVOT lightcurves with the XRT canonical lightcurve model.

\subsection {The early UVOT afterglow}
There are several physical mechanisms and geometric scenarios that may produce a rise in the early optical afterglow. 
In this section the following mechanisms and scenarios shall be discussed: (i) passage of the peak synchrotron 
frequency of the forward shock $\nu_{m,f}$, through the observing band, (ii) a reverse shock, (iii) decreasing 
extinction with time, (iv) the onset of the forward shock in the cases of an isotropic outflow or a jet 
viewed in a region of uniform energy density, (v) the rise produced by an off-axis jet, which may be 
structured, and finally, (vi) a two component outflow.

\subsubsection{Passage of the synchrotron frequency, $\nu_{m,f}$}
\begin{figure*}
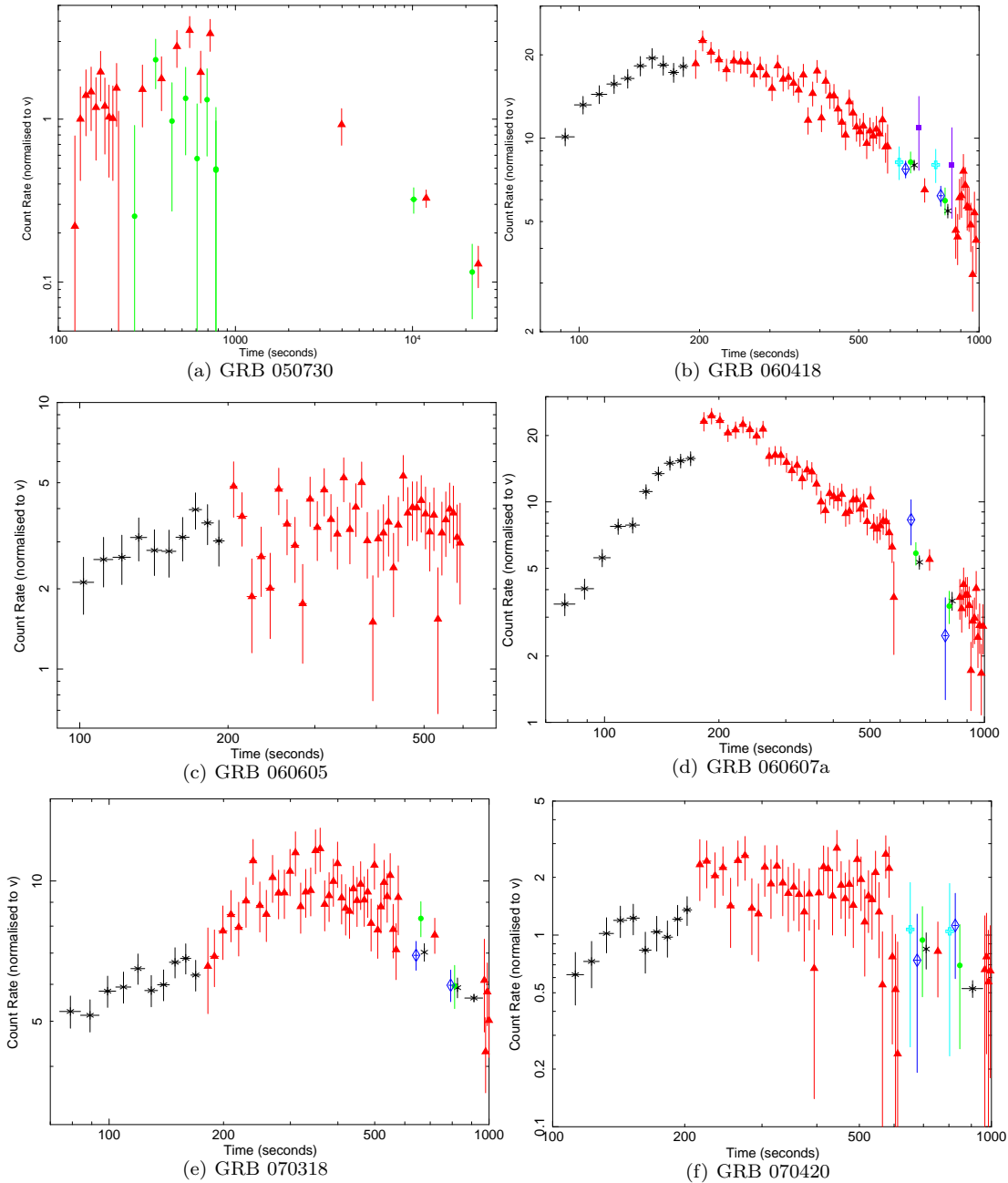

\begin{center}
\subfigure[GRB~050730]{\label{fig:edge-d}\includegraphics[scale=0.3,angle=-90] {./Images/GRB050730.cps}}
\subfigure[GRB~060418]{\label{fig:edge-d}\includegraphics[scale=0.3,angle=-90] {./Images/GRB060418.cps}}
\subfigure[GRB~060605]{\label{fig:edge-d}\includegraphics[scale=0.3,angle=-90] {./Images/GRB060605.cps}}
\subfigure[GRB~060607a]{\label{fig:edge-d}\includegraphics[scale=0.3,angle=-90]{./Images/GRB060607a.cps}}
\subfigure[GRB~070318]{\label{fig:edge-d}\includegraphics[scale=0.3,angle=-90] {./Images/GRB070318.cps}}
\subfigure[GRB~070420]{\label{fig:edge-d}\includegraphics[scale=0.3,angle=-90] {./Images/GRB070420.cps}}
\end{center}
\caption{The early normalized lightcurves for the 6 GRBs with observed rises. GRB~050730 is the 
only GRB observed to rise that was not viewed with the white filter. Key: black star = white, 
red filled triangle = v, green circle = b, blue diamond = u, light blue cross = uvw1, 
pink hexagon = uvm2, purple square =uvw2 }
\label{early_afterglows}
\end{figure*}

The first mechanism, the passage of the peak frequency of the forward shock which moves with time as 
$\nu_{m,f}\propto t^{-3/2}$, through the observing band is expected to produce a chromatic peak in 
the optical lightcurve evolving from the shortest wavelengths through to the longest wavelengths. 
For 5 of the lightcurves with a peak, the UVOT observed the majority of the rise and the peak during 
the two finding chart exposures observed in white and v (see Fig. \ref{early_afterglows}). 
If the peak were due to the passage of the synchrotron frequency, a stepped decrease in flux 
would be observed in the normalized lightcurves at the transition between the white and v 
observations, which has not been observed in any of these GRBs. For the 6th GRB with an observed 
peak, GRB~050730, the rise was observed in the v and b filters. If the passage of $\nu_{m,f}$ was 
the cause of this rise, the afterglow would appear to be brighter in the b filter than in the v filter during the rise, 
and begin to decay earlier. However, the afterglow is not brighter in the b filter than in the v filter during the rise. 
Therefore, the passage of $\nu_{m,f}$ through the optical band is not responsible for any of the peaks observed in 
these optical afterglows.

During the passage of $\nu_m$ from the shortest wavelengths through to the longest wavelengths, the spectrum 
of the optical afterglow changes (assuming slow cooling) from $\nu^{1/3}$, for $\nu<\nu_m$ to 
$\nu^{-(p-1)/2}$, for $\nu_m<\nu<\nu_c$ \citep{sari98}. As $\nu_m$ passes from high frequencies to lower frequencies, 
there will be a change in colour. The colour change between the white and v filters can be calculated using the 
central wavelengths of the white and v filters, given in \cite{poole}, converted in to frequency: 
$\nu_{\rm white}=8.64\times10^{14}$ Hz, $\nu_{\rm v}=5.49\times10^{14}$Hz and assuming $p=2.3$ where $p$ is 
the electron energy index. The colour change between white and v as $\nu_m$ 
moves from above the white frequency to below the v frequency is 0.48 magnitudes. In the lightcurves of Fig. 
\ref{early_afterglows}, which have been normalized using the late time data, the colour difference between the 
white and the v filter after the peak is zero. None of the 5 lightcurves, in Fig. \ref{early_afterglows}, with 
early white and v observations, show such a large offset between the lightcurves in two filters during the rise. 
Furthermore, the equation for $\nu_{m,f}$ in a constant density medium, as given by \cite{zhang05} as:  
\begin {equation}\nu_{m,f}\,=(6\times10^{15}Hz)(1+z)^{1/2}E_{52}^{1/2}\epsilon_e^2\epsilon_B^{1/2}(t/1 day)^{-3/2}\end{equation}
where $E_{52}=10^{52}E$ is the isotropic energy in units of $10^{52}$ ergs, $\epsilon_e$ is the fraction of energy in the electrons, 
$\epsilon_B$ is the fraction of energy in the magnetic field, $z$ is the redshift and $t$ is the time. 
Assuming $\epsilon_e$ and $\epsilon_B$ are not evolving with time and given the time $t_1$ at which $\nu_{m,f}$ is 
at a given frequency $\nu_1$, the time $t_2$ at which $\nu_{m,f}$ is at the frequency $\nu_2$ is given by: 
$t_2=t_1(\nu_2/\nu_1)^{(-2/3)}$. Using the central wavelengths of the white and v filters, the v filter should peak 
1.35$\times$ later than the white filter. There does not appear to be any time difference between the peak in the white and v 
filters in the 5 lightcurves shown in \ref{early_afterglows}, therefore, the passage of $\nu_m$ through the optical band is 
not the cause the rise in the optical band.

For $\nu_{m,f}$, to have passed below the v filter (5402{\AA}) by the time the UVOT has begun observations ($t\sim100$s), using 
$\langle E_{k,iso}\rangle=1.5\times10^{54}$ erg and $\langle z\rangle=2.21$, then 
$\epsilon_e^2\epsilon_B^{1/2}<1.7\times10^{-5}$. The values $\epsilon_e$ and $\epsilon_B$ 
provided in \cite{pan02} for 10 GRBs give values for $\epsilon_e^2\epsilon_B^{1/2}$ ranging from 
$3\times10^{-3}$ to $2\times10^{-7}$, suggesting that $1.7\times10^{-5}$ is consistent with values found for other GRBs.

\subsubsection{The reverse shock}
Considering a constant density medium, there are two main cases of the reverse shock that depend on 
the position of the peak synchrotron frequency of the reverse shock $\nu_{m,r}$ relative to the 
optical band $\nu_{opt}$. If $\nu_{m,r}<\nu_{opt}$ then the lightcurve produced by the reverse shock is 
expected to decay immediately after the peak with $\alpha=(3p+1)/4$ \citep{zhang03}, where $p$ is the 
electron energy index. The value of $p$ is typically taken to lie between 2 and 3, therefore $\alpha$ 
is expected to range between $\alpha=-1.75$ for $p=2$ to $\alpha=-2.5$ for $p=3$. Within the sample, GRB~050726, GRB~061007 and 
GRB~070529 are the only GRBs with decays before 500s that are consistent at 2$\sigma$ with the slowest 
expected reverse shock decay index $\alpha=-1.75$; GRB~050726 has a decay of $\alpha=-2.67 \pm 0.80$, 
GRB~061007 decays with $\alpha=-1.63 \pm 0.11$ and GRB~070529 decays with $\alpha=-1.64\pm 0.14$. The other lightcurves 
in the sample are shallower than the reverse shock prediction with $>2\sigma$ confidence. 
The second case arises if $\nu_{m,r}>\nu_{opt}$, then immediately after the peak there is an 
intermediate stage where $\alpha\sim~-0.5$, which is followed by $\alpha=(3p+1)/4$. In the sample, 
there are 7 GRBs that before 500s have temporal indices consistent with $\alpha=- 0.5$ at $2\sigma$ confidence.
However, of these GRBs only GRB~060223a has a decay after 500s ($\alpha=-1.53 \pm 0.60$) that is consistent 
at the $2\sigma$ level with the slowest expected reverse shock decay index $\alpha=-1.75$. 

The reverse shock in a wind medium is expected to produce a much steeper decay immediately after 
the peak with $\alpha\sim-3.5$ \citep{kobay04}. Only GRB~050726, has a value of $\alpha_{<500s}$, 
which is consistent to within 2$\sigma$ confidence. All other GRBs are inconsistent at $>5\sigma$.

The inconsistency of the temporal indices of the GRBs in this sample to the temporal decay expected during a 
reverse shock for both types of medium implies that the reverse shock is not the main contributor to the 
optical emission at early times, and therefore is not responsible for the rise. Still the reverse shock 
is expected to occur for all relativistic outflows that interact with the external medium. For a number 
of GRBs, the reverse shock may not be observed as it can be suppressed by high levels of magnetisation 
in the outflow \citep{zhang05,giannios08} or if the forward and reverse shock have comparable energy, 
the sharp decay in the reverse shock may be masked by the flux produced by the forward shock \citep{mcmahon06,mundell07}. 

\subsubsection{Dust Destruction}
If there are high levels of extinction at the beginning of the afterglow \citep[][and references within]{klotz08}, the 
lightcurve produced would be dim and reddened at the beginning. As the dust is destroyed by the radiation, a chromatic peak 
would be observed as the afterglow brightens and becomes less red. The bluer filters would be expected to rise more 
steeply when compared to the red filters as would the white filter, because the sensitivity of this filter is skewed to the blue. 
However, the amount of dust destroyed is highly dependent on the environment of the GRB, in particular to the density and the 
size of the region surrounding the progenitor, and simulations suggest that most of the dust destruction is expected to occur 
within the first few tens of seconds after the trigger \citep{perna02}. Therefore, it is unlikely that the UVOT is observing 
the afterglow while dust destruction is occurring.

However, as the duration of dust destruction and the quantity of dust destroyed are only theoretical predictions, we must rule 
out dust destruction using observations. Therefore, the 6 UVOT lightcurves with a rise were examined 
to see if the bluer filters, including white, rise more steeply when compared to the v filter. The UVOT observed five of the 
lightcurves in white and v (the reddest UVOT filter) during the rise, see Fig. \ref{early_afterglows}. GRB~060607a has the only 
lightcurve where there appears to be a significant excess in v compared with the white filter. However, the H band lightcurve 
given in \cite{mol06} rises at the same rate as the UVOT lightcurve, which was observing in white during the rise. If dust 
destruction was the cause of the rise, the H band would be expected to rise less steeply than the UVOT lightcurve. The 6th 
rising GRB, GRB~050730, was observed with the v and b filters during the rise and peak. If the peak in this case were due 
to dust destruction, an excess in v compared to b would be observed. However, the lightcurves of the v and b filters are 
consistent within 1$\sigma$ errors. Therefore, there is no evidence to suggest that dust destruction is the cause of the 
rise for this GRB or for any of the GRBs in this sample.

\subsubsection{Start of the forward shock}
At the start of the forward shock, a rise will be observed in the lightcurves as the jet ploughs into the external medium \citep{sari99}. 
The lightcurves for an observer viewing within a uniform jet, or within a cone of uniform energy density will be the same as those 
observed within an isotropic outflow \cite[and references within]{granot02}. The temporal index of the rise will vary according to 
the thickness of the shell and the density of the external medium. Assuming the synchrotron self absorption frequency 
$\nu_{a}<\nu_{opt}$, then the thickness of the shell and the density profile of the external medium affect the rate at 
which the lightcurve rises. For the thick shell case, the temporal index is $\alpha = 1$ for a constant density medium, 
or $\alpha=1/9$ for a wind medium. For the thin shell case, in a constant density medium the temporal index is either 
$\alpha=2$ for $\nu_{c}<\nu_{opt}$, or $\alpha=3$ for $\nu_{c}>\nu_{opt}$. Lastly, the temporal index is $\alpha=0.5$ 
if the shell is thin and in a wind medium \citep{pan08}. 

Given the peak time, the Lorentz factor $\Gamma$ of the shell at the moment of the peak for a constant density medium, can be derived using the
 following equation \citep{mol06,sari97}:
 \begin {equation}\Gamma(t_{peak})=\left(\frac{3E(1+z)^3}{32\pi nm_pc^5\eta (t_{peak}/100{\rm s})^3}\right)^{1/8}\end{equation}
where $t_{peak}$ is the peak time, $\eta$ is the radiative efficiency and $n$ is the density of the medium. 
Here we will assume $\eta=0.2$ and $n=1$ cm$^{-3}$. However, changing the values of $\eta$ and $n$ has a minor 
effect on the final values of $\Gamma(t_{peak})$ as the dependence of $\Gamma(t_{peak})$ on these parameters is small: 
$\Gamma(t_{peak})\propto (\eta n)^{-1/8}$. For each GRB, the k-corrected energy, given in Table \ref{redshift}, was 
used in the equation above to determine $\Gamma(t_{peak})$ and the resulting values of $\Gamma(t_{peak})$ for the 
individual GRBs can be found in Table \ref{redshift}. The mean of the $\Gamma(t_{peak})$ for the GRBs with a 
peak in their lightcurve is $\langle{\Gamma(t_{peak})}\rangle\sim 180$, which is consistent with the expectation 
that the initial Lorentz factor of the jet $\Gamma_0$, where $\Gamma_0\sim2\Gamma(t_{peak})$ \citep{pan00, mes06}, 
of GRBs is $>100$ \citep{fen93}. For the GRBs where only an upper limit to their peak time is known, the mean 
value for $\Gamma(t_{peak})$ is a lower limit, $\langle{\Gamma(t_{peak})}\rangle >230$. This suggests that the GRBs 
with observed peaks typically have lower Lorentz factors than the GRBs with upper limits to their peak times.

Using the derived values of $\Gamma(t_{peak})$, it is possible to deduce two more quantities: the isotropic-equivalent mass 
of the baryonic load $M_{fb}=E/(\Gamma_0 c^2)$, and the deceleration radius $R_{dec}\simeq2ct_{peak}\Gamma(t_{peak})^2/(1+z)$ \citep{mol06}. 
These quantities were determined for each GRB and the results are given in Table \ref{energies}. The mean mass of the 
baryonic load and the mean deceleration radius for the GRBs with an observed rise are $\langle{M_{fb}}\rangle=3.8\times10^{-3}M_{\odot}$ 
and $\langle{R_{dec}}\rangle=2.1\times10^{17}$cm and for the GRBs without an observed rise the quantities are 
$\langle{M_{fb}}\rangle<1.1\times 10^{-3}M_{\odot}$ and $\langle{R_{dec}}\rangle<1.2\times 10^{17}$cm. The deceleration radii are 
in agreement with $R_{dec}\sim10^{16}\rm cm$ as predicted by theory \citep{rees92}. Therefore the forward shock is consistent with our observations.

\subsubsection{Off-axis and structured outflows}
A rise may be produced in the lightcurve if the observer's viewing angle is $\theta_{obs}>\theta$, where 
$\theta_{obs}$ is the observers viewing angle and $\theta$ is the half opening angle 
of the jet. In the case of a uniform jet, the ejecta are released into a cone of angle $\theta$ and 
due to relativistic effects, the emission in the jet is beamed as $\Gamma^{-1}$. If $\theta_{obs}>\theta$, 
then the emission is strongly beamed away from the observer, but as  $\Gamma$ decreases, the emission 
entering the line of sight increases and the observed lightcurve will rise. The 
lightcurve will peak when $\Gamma\sim(\theta_{obs}-\theta)^{-1}$ \citep{granot02}. Larger observing angles will view a later peak 
and the peak magnitude will be lower \citep{granot05}. A structured jet, in which the energy per solid 
angle decreases around a uniform core of angular size $\theta_c$, viewed off-axis ($\theta_{obs}>\theta_c$) 
can produce a rise in the optical afterglow, where the behaviour of the rise may vary depending on the 
distribution of energy around the core. In this case, the more diffuse the energy per solid angle the slower 
the rise and the later the peak time \citep{pan08,granot02}. The peak time of the lightcurve is also dependent 
on the viewing angle of the jet and the peak will occur when $\Gamma\sim(\theta_{obs}-\theta_c)^{-1}$.

\subsubsection{Two Component Outflows}
A two component outflow consists of a narrow jet surrounded by a wide jet. The narrow component will be 
denoted by a subscript $n$ and the wide component will be denoted by a subscript $w$. Both components 
move at relativistic speeds, but the narrow jet will have a larger $\Gamma$ (i.e $\Gamma_w<\Gamma_n$) and the 
wide component will have a larger half opening angle than the inner narrow jet (i.e $\theta_w<\theta_n$). The 
optical emission is expected to be produced either within the wide component \citep{depas08,oates07} or 
more traditionally from both the narrow and wide components \citep{peng05,huang04}. 

In the case where the optical emission is only produced by the wide component, the rise observed will occur as in Section 
5.1.4, provided that the jet is viewed within $1.5\theta_w$, as has been demonstrated by \cite{granot02}. 
At angles larger than $1.5\theta_w$, the peak of the optical lightcurve will occur when $\Gamma\sim(\theta_{obs}-\theta_{w})^{-1}$.
This model of the two component outflow could produce the rises observed in this sample.

In the case where the optical emission is produced in both the narrow and the wide components, the dominance of emission 
from one component over the other, depends mostly on the energy within each component and on the viewing angle of the 
observer (for a more detailed description see \cite{peng05}). However, in this case as the emission is produced in 
both components, it is likely that on-axis and off-axis viewers will observe two peaks. This effect is not seen 
in the UVOT lightcurves in this paper and therefore, the jet is unlikely to have two components where both produce 
optical/UV emission.

\subsection{The Late UVOT afterglow}

A correlation has been observed between the observed magnitude at 400s and the decay after 500s. However, there is 
no significant evidence from a Spearman Rank test performed between $\alpha_{>150s,rest}$ and the luminosity in the 
rest frame at 100s, for a similar correlation in the rest frame. For an off-axis jet, a correlation is expected between the 
luminosity and the decay of the lightcurve when the viewing angle is changed, with fainter, shallower afterglows 
having a larger viewing angle \citep{pan08}. As this data do not show a strong correlation of this type, this suggests 
that the outflows are viewed within the half-opening angle $\theta$ or within a core of uniform energy density $\theta_c$. 
This supports the idea that the start of the forward shock produces the rises observed in the UVOT lightcurves. However, 
this does not give an indication of the geometry of the jet which may be uniform, comprise of two components, or be structured. 

\subsection {UVOT afterglow luminosity}

The luminosity distribution found in this work shows no evidence for bimodality, which is in contradiction 
with the results of \cite{nard06,nard08}, \cite{liang} and \cite{kann07}, who all claim a bimodal distribution 
within their samples. However, the lack of evidence for bimodality within the distributions in this 
sample is consistent with the work of \cite{cenko08} and \cite{mel08}. \cite{cenko08} present a luminosity distribution at 1000s 
in the rest frame from a sample of 17 GRBs, with known redshift, observed with the Palomar 60 inch telescope. Their distribution 
shows no evidence for bimodality. \cite{mel08} produce 3 luminosity distributions for 16 optical afterglows observed with the Liverpool and 
Faulkes telescopes. They find a single peaked luminosity distribution for three rest frame epochs: 10mins, 0.5 days and 1 day, which are well 
fit by a log-normal function. \cite{mel08} do not correct their lightcurves for host extinction, but as discussed 
in this paper the correction for host extinction appears to have minimal effect on the luminosity distribution. 
Therefore, it is possible to compare the distributions of \cite{mel08} with the distributions produced with this sample.  

\subsection {Comparison of the XRT and UVOT canonical lightcurves}
There are three segments which are usually found in X-ray lightcurves within the first $\sim10^{5}$s \citep{zhang05, nousek}. 
The first segment is a fast, early decay with $-5<\alpha_{X1}<-3$, typically ending within 100s-1000s after the trigger. 
The fast decay is thought to be caused by the tail of the prompt emission \citep{zhang05, nousek}. The second segment is 
shallow with $-1.0<\alpha_{X2}<-0.5$ \citep[although this range should now be considered as $-1.0<\alpha_{X2}<0.0$,][]{liang07}, 
ceasing between 1000s and 10000s and is attributed to energy injection \citep{zhang05, nousek}. The third segment decays 
as $-1.5<\alpha_{X3}<-1$ and is expected to occur at the end of energy injection \citep{zhang05,nousek}.

The range in temporal index of the optical lightcurve before 500s, taken as the mean plus or minus the dispersion, is $-1.17<\alpha<0.21$. 
This range is clearly inconsistent with the first segment of the XRT canonical lightcurve. The range of temporal index before 
500s is most similar to the range of the second segment of the XRT canonical model. However, none of the XRT canonical lightcurve segments 
indicate rising behaviour for the X-ray lightcurves. Applying the theoretical interpretations for the individual segments 
of the XRT canonical lightcurve, provided by \cite{zhang05} and \cite{nousek}, to the UVOT canonical lightcurve suggests that before 500s, the emission 
producing the optical lightcurves is from the forward shock and that a number of the UVOT afterglows during this period 
are energy injected. The lack of corresponding rising behaviour in the X-ray lightcurves, presuming an achromatic 
rise ($\nu_m<\nu_{opt}$), suggests that the rise of the forward shock is masked in the X-rays, possibly by the 
contribution of the prompt emission. This is consistent with the model of \cite{willingale07} who suggest that the steep 
and shallow decays of the X-ray lightcurves are dominated by the prompt emission and afterglow emission respectfully, 
and that they do not observe the rise of the afterglow emission as this is masked by the prompt emission.

In Section 4, the lightcurves after 500s were fitted with power-laws and broken power-laws. The GRBs which were best 
fitted by broken power-laws are discussed separately. For the optical lightcurves that decay as power-laws, 
the range in decay is $-1.22<\alpha<-0.52$. Like the range in decay before 500s, the range in decay after 500s is most
similar to the decay range of the second segment of the XRT canonical model. Assuming the same reasoning as for 
the XRT segments, this suggests that after 500s the optical lightcurves are consistent with emission from the forward 
shock and most of these are energy injected.

For the four GRBs that are best fit with a broken power-law after 500s, the range in temporal decay before the break is 
$-0.74<\alpha<-0.46$, which is consistent with the range given for the second segment of the canonical XRT model. This 
suggests that before the break the optical lightcurves are energy injected. The range in the temporal decay after the 
break is $-1.72<\alpha<-1.34$, which is consistent with the third decay of the XRT canonical lightcurve, which has 
been suggested to be the decay following the end of energy ejection.

\section{Conclusions}
In this paper we systematically reduced and analyzed a sample of 27 GRBs, which met a strict set of selection criteria.
We note that the temporal behaviour of the optical afterglows in the sample is varied, with the greatest 
variation occurring in the early phase of the lightcurves: before 500s the lightcurves may rise or decay. 
The mean for $\alpha_{<500s}$ is $-0.48^{+0.15}_{-0.19}$ with a dispersion of $0.69^{+0.19}_{-0.06}$. However, after 500s, 
all the lightcurves decay. The lightcurves were fitted with power-laws and broken power-laws. A broken power-law was 
deemed to be an improvement, if the $\chi^2/D.O.F$ decreased and the probability of chance improvement was small ($<1\%$) 
and in 5 cases a broken power-law was considered a better fit. The mean decay index after 500s, when including only 
those that decay with a single power-law, is $-0.87^{0.10}_{-0.09}$ with a dispersion of $0.35^{+010}_{-0.04}$. There 
is a correlation at $99.8\%$ probability, between the magnitude at 400s and the temporal decay after 500s, with the 
brightest optical afterglows decaying the fastest. 

We investigated the cause of the rising behaviour in the early afterglow and discussed the following physical 
mechanisms and geometric scenarios: the passage of the synchrotron frequency $\nu_m$, reverse shock, dust 
destruction, the start of the forward shock, the viewing angle of a (possibly structured) jet and a two component outflow. The rise 
in the optical lightcurves may be attributed to either the start of the forward shock, or to an off-axis 
viewing angle where the observer sees an increasing amount of emission as the Lorentz factor of the jet 
decreases. We also investigated the correlation between magnitude and decay after 500s. We determined that a 
correlation observed between the magnitude at 400s and the decay after 500s is only weakly dependent on redshift. 
A Spearman rank test performed between the luminosities at 100s and the decay after 150s in the rest frame did 
not reveal a significant correlation. However, a luminosity-decay correlation would be expected for jets viewed off-axis, 
where the more off-axis the jet the fainter and shallower the lightcurve. We do not observe a strong correlation 
of this type, suggesting that the optical lightcurves are produced by jets viewed on-axis and that the rise observed 
in the optical lightcurves is caused by the start of the forward shock.

We produced luminosity lightcurves for the 21 GRBs in the sample with known redshift. The luminosity 
lightcurves were produced at a common wavelength of 1600{\AA} in the rest frame. We find that the logarithmic distribution of 
the luminosities at three rest frame epochs: 100s, 1000s and 10~ks do not show evidence for bimodality. Correcting 
the lightcurves for the host extinction increases the mean luminosities of the distributions, but does not 
considerably alter their appearance and the change in standard deviations of the logarithmic luminosity 
distributions is no greater than 0.08 for any of the three epochs. The lack of evidence for bimodality 
is consistent with the findings of \cite{mel08} and \cite{cenko08}.

Finally, we compared the temporal behaviour of the optical afterglows in this sample with the XRT canonical model.
We have found that the temporal indices before 500s and the temporal indices of the lightcurves after 
500s are most consistent with the the shallow decaying segment of the XRT canonical model. \cite{nousek} and 
\cite{zhang05} suggest that the shallow decay segment of the XRT canonical model is energy injected. This 
would suggest that the optical lightcurves are energy injected as well. The lack of rises observed in X-ray 
afterglows could be due to the prompt emission masking them.

\section{Acknowledgments}
This research has made use of data obtained from the High Energy Astrophysics Science Archive Research Center 
(HEASARC) and the Leicester Database and Archive Service (LEDAS), provided by NASA's Goddard Space Flight Center 
and the Department of Physics and Astronomy, Leicester University, UK, respectively. SRO acknowledges the support 
of an STFC Studentship. SZ thanks STFC for its support through an STFC Advanced Fellowship. %


\bibliographystyle{mn2e}   
\bibliography{OatesSR} 


\IfFileExists{\jobname.bbl}{}
 {\typeout{}
  \typeout{******************************************}
  \typeout{** Please run "bibtex \jobname" to optain}
  \typeout{** the bibliography and then re-run LaTeX}
  \typeout{** twice to fix the references!}
  \typeout{******************************************}
  \typeout{}
 }

\end{document}